\documentclass[structabstract]{aa}
\usepackage{natbib}
\usepackage{lscape}
\usepackage{rotating}
\usepackage{graphicx}
\usepackage{amsmath}
\usepackage{enumerate}
\usepackage{txfonts}
\usepackage{color}
\usepackage{hyperref}

\newcommand{\Msun}{M_{\odot}}
\newcommand{\Rsun}{R_{\odot}}

\newcommand{\Ebind}{E_{\rm bind}}

\begin{document}

  \title{Common-envelope ejection in massive binary stars}
  \subtitle{Implications for the progenitors of GW150914 and GW151226}
  
   \author{M.~U.~Kruckow
          \inst{1}\fnmsep\thanks{e-mail: mkruckow@astro.uni-bonn.de},
          T.~M. Tauris\inst{2,1},
          N. Langer\inst{1},
          D. Sz\'{e}csi\inst{1,3},
          P. Marchant\inst{1}
           \and
          Ph. Podsiadlowski\inst{4,1}
          }

   \authorrunning{Kruckow~et~al.}
   \titlerunning{CE ejection and GW150914}

   \institute{Argelander-Institut f\"ur Astronomie, Universit\"at Bonn, Auf
              dem H\"ugel 71, 53121 Bonn, Germany
         \and
              Max-Planck-Institut f\"ur Radioastronomie, 
              Auf dem H\"ugel 69, 53121 Bonn, Germany 
         \and 
              Astronomical Institute of the Czech Academy of Sciences,
              Ond\v{r}ejov, Czech Republic 
         \and
              Department of Astronomy, Oxford University, Oxford OX1~3RH, UK 
              }

  \date{Received July 28, 2016; accepted October 14, 2016}

 
  \abstract 
{The recently detected gravitational wave signals (GW150914 and GW151226) of the merger event of a pair of relatively massive
 stellar-mass black holes (BHs) calls for an investigation of the formation of such progenitor systems in general.}  
{We analyse the common-envelope (CE) stage of the traditional formation channel in binaries where the first-formed compact object undergoes an
 in-spiral inside the envelope of its evolved companion star and ejects the envelope in this process.}  
{We calculated envelope binding energies of donor stars with initial masses between 4 and $115\;M_{\odot}$ for metallicities of 
 $Z=Z_{\rm Milky\,Way}\simeq Z_{\odot}/2$ and $Z=Z_{\odot}/50$, and derived minimum masses of in-spiralling objects needed to eject these envelopes.}  
{In addition to producing double white dwarf and double neutron star binaries, CE evolution may
also produce
 massive BH-BH systems with individual BH component masses of
up to $\sim\!50-60\;M_{\odot}$, in particular for donor stars evolved to giants
 beyond the Hertzsprung gap. However, the physics of envelope ejection of massive stars remains uncertain. We discuss the
 applicability of the energy-budget formalism, the location of the bifurcation point, the recombination energy, and the accretion
 energy during in-spiral as possible energy sources, and also comment on the effect of inflated helium cores.}  
{Massive stars in a wide range of metallicities and with initial masses of up to at least $115\;M_{\odot}$ may shed their
 envelopes and survive CE evolution, depending on their initial orbital parameters, similarly to the situation for intermediate-
and low-mass stars with degenerate cores. In addition to being dependent on stellar radius, the envelope binding energies and
 $\lambda$-values also depend on the applied convective core-overshooting parameter, whereas these structure parameters are
 basically independent of metallicity for stars with initial masses below $60\;M_{\odot}$. Metal-rich stars $\ga 60\;M_{\odot}$ become
 luminous blue variables and do not evolve to reach the red giant stage. We conclude that based on stellar structure calculations, 
 and in the view of the usual simple energy budget analysis, events like GW150914 and GW151226 might be produced by the CE channel. 
 Calculations of post-CE orbital separations, however, and thus the estimated LIGO detection rates, remain highly uncertain.}


   \maketitle

\section{Introduction}\label{sec:intro}
The majority of all massive stars are found in close binaries that will eventually interact through mass transfer during their stellar lifetimes \citep{sdd+12}. 
This sometimes leads to the formation of compact stellar X-ray sources \citep[e.g.][]{tv06} and, in some cases, to the eventual production of a 
pair of compact objects merging within a Hubble time. The evolution of massive single stars \citep[e.g.][]{hfw+03} has been investigated for many years 
and is still far from being well understood. The evolution of massive (interacting) binary stars is even more complex and can be significantly different \citep{pjh92,lan12}.

Common-envelope (CE) evolution is thought to play a key role in the formation of many close-orbit binaries containing two compact objects,
that is, white dwarfs (WDs), neutron stars (NSs), or black holes (BHs). Given their current small orbital separation (often much
smaller that the radii of their progenitor stars), a binary interaction process must have been at work to reduce the orbital
energy and angular momentum significantly. CE evolution is a good candidate for such a process since it is accompanied by 
a drag-force, arising from the motion of the in-spiralling object through the envelope of its companion star, which leads to
dissipation of orbital angular momentum and deposition of orbital energy in the envelope. Hence, the global outcome of a CE phase is
a reduced binary separation and ejected envelope, unless the system coalesces. The final post-CE separation, however,
is difficult to predict as a result of our poor understanding of the complex physical processes involved in envelope ejection. The huge
ranges in both length scales and timescales make hydrodynamical simulations troublesome. For general reviews on CE evolution, 
see for instance \citet{il93,tas00,pod01,ijc+13}.

There is strong evidence of past orbital shrinkage (i.e. similar to the expected outcome of a CE phase) in a number of observed close binary
pulsars and WD pairs with orbital periods of a few hours or less. Examples include PSR~1913+16 \citep{ht75}, PSR~J0737$-$3039
\citep{bdp+03}, CSS~41177 \citep{bmp+14}, and J0651+2844 \citep{bkh+11}. These systems are tight enough that gravitational-wave 
radiation will bring the two compact objects (e.g. NS-NS or WD-WD binary) into contact within a Gyr, which in some cases leads to a merger event.

Similarly, the recent, and first, gravitational wave detection GW150914 \citep{aaa+16} of the merger event of two relatively massive stellar-mass BHs ($36+29\;M_{\odot}$) 
raises interesting questions about its origin. This system has also been suggested to form through CE evolution \citep{bhbo16}.
However, for massive stars there are other formations channels in which a binary system may evolve to become a tight pair of BHs.
Three main formation channels to produce such a BH-BH pair are
\begin{itemize}
 \item[i)] the CE formation channel (i.e. traditional channel), 
 \item[ii)] the chemically homogeneous evolution (CHE) channel with or without a massive overcontact binary (MOB), and 
 \item[iii)] the dynamical channel in dense stellar environments. 
\end{itemize}

i) The CE formation channel for BHs is similar to that which is believed to produce tight double NS systems \citep[e.g.][and references therein]{tv06}.
In this scenario, the systems always enter a CE phase following the high-mass X-ray binary (HMXB) stage,
during which the O-type star becomes a red supergiant and captures its BH companion.
There are many uncertainties, however, involved in calculations of the in-spiral and the subsequent CE ejection. The evolution is often 
tidally unstable, and the angular momentum transfer, dissipation of orbital energy, and structural changes of the donor star take place on very 
short timescales \citep[$<10^3\;{\rm yr}$,][]{pod01}. A complete study of the problem requires detailed multi-dimensional hydrodynamical calculations,
although early studies in this direction have difficulties ejecting the envelope and securing deep in-spiral \citep{tas00,pdf+12,rt12,nil14,orps16}.
The calculations along this route are therefore highly uncertain owing to our current poor knowledge of CE physics \citep{ijc+13}.
As a consequence, the predicted detection rate of BH-BH mergers from the CE channel is uncertain
by several orders of magnitude \citep{aaa+10}, also partly as
a result of the unknown amount of (asymmetric) mass loss
that is associated with a possible supernova explosion (i.e. imparted momentum kick) when a BH is formed. 
Examples of population syntheses investigations of BH-BH binaries following the traditional channel with a CE scenario include \citet{bkb02,vt03,bkr+08,dbf+12,mv14,bhbo16,es16}.

ii) The two other formation channels of BH-BH binaries avoid the CE phase altogether. In the CHE scenario for binaries \citep{dcl+09,md16,dm16},
the stars avoid the usual strong post-main sequence expansion as a result of effective mixing  
enforced through the rapidly rotating stars through tidal interactions. Hence, this works only for massive stars
at low metallicity where strong angular-momentum loss due to stellar winds can be avoided.
\citet{mlp+16} presented the first detailed CHE models leading to the formation of BH-BH systems
and demonstrated that MOB systems are particularly suited for this channel, enabling formation of very massive stellar-mass BH-BH mergers, 
that is in agreement with the detection of GW150914. Lower-mass BH-BH mergers like GW151226 ($14+8\;M_{\odot}$), however, cannot be formed from this scenario.

iii) Finally, the dynamical formation channel \citep[e.g.][]{sh93,pm00,bbk10,rcr16} produces BH-BH mergers through encounter interactions in dense stellar clusters
and thereby circumvents the need for mass transfer and CE evolution. In analogy to the other production channels mentioned above, the rate of
BH-BH mergers from the dynamical formation channel is also difficult to constrain. Recent studies predict that this channel probably
accounts for less than about 10\% of all BH-BH mergers \citep[e.g.][]{rcr16}. 

In this paper, we investigate the prospects of envelope ejection from massive stars during the CE stage with an in-spiralling compact object, following the CE formation channel.
In Sect.~\ref{sec:CE_ejection} we introduce the CE ejection criterion based on energy budget considerations. In Sect.~\ref{sec:results}
we present our calculated envelope binding energies and so-called $\lambda$-values of donor stars with initial zero-age main-sequence (ZAMS) masses between 4 and $115\;M_{\odot}$ 
for different metallicities (Milky~Way-like: $Z_{\rm MW}\approx Z_{\odot}/2$, and IZw18-like: $Z=Z_{\odot}/50$), and derive minimum masses of in-spiralling compact objects 
(or non-degenerate stars) needed to eject these envelopes
based on simple energy considerations. In addition, we analyse the stellar structure of pre-CE donor stars, with the aim of better understanding the location of the core boundary.
A general discussion of our results is given in Sect.~\ref{sec:discussions}, where we also revisit the question of
locating the bifurcation point of envelope ejection, debate the possibility of additional energy input from liberated recombination energy or accretion during in-spiral, 
and comment on the effect of inflated helium cores. Finally, we briefly discuss our results in relation to the LIGO merger events GW150914 and GW151226 in Sect.~\ref{sec:ligo}, 
before summarising our conclusions in Sect.~\ref{sec:conclusions}.

\section{Criterion for common-envelope ejection}\label{sec:CE_ejection}
The central problem in question is whether a massive binary will survive a CE evolution or result in an early merger event without ever forming a
BH-BH system. Whether the donor star envelope is ejected successfully depends on the binding energy of the envelope, the available energy resources to expel it, 
and the ejection mechanism.

Following the ($\alpha,\lambda$)-formalism introduced by \citet{web84} and \citet{dek90}, we can write the 
criterion for successful envelope ejection as $E_{\rm bind}\le \alpha\,\Delta E_{\rm orb}$, where $\alpha$
is the efficiency of converting released orbital energy into kinetic energy that provides the outward ejection
of the envelope. The total binding energy (gravitational plus internal thermodynamic contributions) of the donor star envelope at onset of the CE is given by\begin{equation}
  E_{\rm bind}= \int _{M_{\rm core}}^{M_{\rm donor}} \left(-\frac{GM(r)}{r}+U\right)\,dm \equiv -\frac{GM_{\rm donor}M_{\rm env}}{\lambda\, R_{\rm donor}} \,
  \label{eq:bindingenergy}
,\end{equation}
where $G$ is the gravitational constant, $M(r)$ is the mass within the radius coordinate $r$ of the donor star with total radius, $R_{\rm donor}$, total mass, 
$M_{\rm donor}$, core mass, $M_{\rm core}$, envelope mass, $M_{\rm env}\equiv M_{\rm donor}-M_{\rm core}$, and $U$ is the specific internal energy \citep{hpe94}. 
Given that $E_{\rm bind}$ is evaluated at the moment the evolved donor star fills its Roche-lobe and initiates 
dynamically unstable mass transfer, leading to formation of the CE, we do not include the gravitational potential from the in-spiralling companion 
when calculating $E_{\rm bind}$ \citep[see e.g.][for alternative descriptions]{pjh92,il93}. 

>From integrations of detailed stellar models, the values of $\lambda$ in Eq.~(\ref{eq:bindingenergy}) can be calculated \citep{dt00,dt01,prh03}
and tabulated (e.g. for use in population synthesis codes). Differences in $\lambda$-values may arise, for example, from
the use of different stellar models, the degree of available recombination energy \citep{ijp15}, enthalpy considerations \citep{ic11,wjl16b}, and, in particular, 
from using different definitions of the core-envelope boundary \citep{td01}. The last problem is discussed in Sect.~\ref{subsec:bif_revisited}.

Since the in-spiral of the companion star often decreases the orbital separation by a factor of $\sim\!100$ or more, we can
approximate the change in orbital energy as
\begin{equation}\label{eq:Eorb}
 \Delta E_{\rm orb} = -\frac{GM_{\rm core}M_{\rm X}}{2\,a_{\rm f}} + \frac{GM_{\rm donor}M_{\rm X}}{2\,a_{\rm i}} \simeq -\frac{GM_{\rm core}M_{\rm X}}{2\,a_{\rm f}} \,
,\end{equation}
where $a_{\rm i}$ and $a_{\rm f}$ denote the initial (pre-CE) and final (post-CE) orbital separation, respectively, and $M_{\rm X}$ is the mass of the in-spiralling object (e.g. $M_{\rm BH}$ for a BH).

It is crucial for our purposes to investigate whether the ability of CE ejection depends on the masses of the two stars; that is
to say, whether it is possible that the CE ejection will work for
producing [8+$8\,M_{\odot}$] BH-BH systems, for instance, but not [30+$30\,M_{\odot}$] BH-BH systems. 
Hydrodynamical simulations and some observational evidence (see Sect.~\ref{subsec:NSNSWDWD}) indicate that the envelope ejection efficiency $\alpha$ does depend 
on the mass of the in-spiralling object, at least in the formation of WD-WD binaries. 
Furthermore, massive stars producing BHs have more tightly bound envelopes, and therefore higher values of $E_{\rm bind}$, than somewhat less massive stars with the same radius. 
The reason for this are the combined effects of a more shallow decline in mass density with radial coordinate and more envelope mass located outside the core boundary 
compared to less massive stars. 
On the other hand, the more massive stars are also able to release more orbital energy from in-spiral to a given final orbital separation.
Therefore, we can consider the ratio $E_{\rm bind}/\Delta E_{\rm orb}$ , which can be rewritten as
\begin{equation}
  \frac{E_{\rm bind}}{\Delta E_{\rm orb}} = \frac{M_{\rm donor}}{M_{\rm X}} \,\frac{2(1-x)}{r_{\rm L}(q')\,x}\,\frac{R_{\rm core}}{\lambda\,R_{\rm donor}},
\label{eq:energy}
\end{equation}
where $x\equiv M_{\rm core}/M_{\rm donor}$, $q'\equiv M_{\rm core}/M_{\rm X}$ and $r_{\rm L}(q\,')\equiv R_{\rm core}/a_{\rm f}$
is the dimensionless Roche-lobe radius \citep{egg83} of the stripped core with mass $M_{\rm core}$ and radius $R_{\rm core}$ (see Sect.~\ref{subsec:He-env} for discussions). 
For the values of $\Delta E_{\rm orb}$, it is assumed that the in-spiral stops just when the remaining core would fill its Roche~lobe.

As a boundary of the remaining core, we take in this study the mass coordinate where the hydrogen abundance $X_H=0.10$  
(see Sects.~\ref{subsec:bifurcation} and \ref{subsec:bif_revisited} for extensive discussions). 

In the above energy formalism, we have assumed a minimum energy requirement, that is, we have neglected any kinetic energy
of the ejected matter and simply assumed that the velocity of the gas is zero at infinity.
In reality, the material may be ejected from the binary with a higher velocity. For example, it has been argued \citep{nil14} 
that the kinetic energy of the ejecta material at infinity might be comparable to the initial binding energy of the envelope of the donor star.
In addition, orbital energy transferred to the ejected material might not have been thermalised \citep{ijc+13}.
The applied energy formalism used to predict the post-CE separation does not take these effects into account, unless a value of $\alpha$ lower than unity is chosen
(see also Sect.~\ref{subsec:alpha}). 

As mentioned above, for successful envelope ejection, it is required that $(E_{\rm bind}/\Delta E_{\rm orb})\le \alpha$ \citep{ls88}.
We now investigate $E_{\rm bind}$ and $\lambda$ for various stellar models and calculate for which values of $M_{\rm X}$ envelope ejection is possible
if the sole energy source to eject the envelope is released orbital energy from the in-spiralling object (which can be a compact object, a star, or a planet).

\section{Results}\label{sec:results}
\begin{figure}[t]
\begin{center}
\begin{minipage}{\columnwidth}
\begingroup
  \fontfamily{Hevetica}%
  \selectfont
  \makeatletter
  \providecommand\color[2][]{%
    \GenericError{(gnuplot) \space\space\space\@spaces}{%
      Package color not loaded in conjunction with
      terminal option `colourtext'%
    }{See the gnuplot documentation for explanation.%
    }{Either use 'blacktext' in gnuplot or load the package
      color.sty in LaTeX.}%
    \renewcommand\color[2][]{}%
  }%
  \providecommand\includegraphics[2][]{%
    \GenericError{(gnuplot) \space\space\space\@spaces}{%
      Package graphicx or graphics not loaded%
    }{See the gnuplot documentation for explanation.%
    }{The gnuplot epslatex terminal needs graphicx.sty or graphics.sty.}%
    \renewcommand\includegraphics[2][]{}%
  }%
  \providecommand\rotatebox[2]{#2}%
  \@ifundefined{ifGPcolor}{%
    \newif\ifGPcolor
    \GPcolortrue
  }{}%
  \@ifundefined{ifGPblacktext}{%
    \newif\ifGPblacktext
    \GPblacktexttrue
  }{}%
  \let\gplgaddtomacro\g@addto@macro
  \gdef\gplbacktext{}%
  \gdef\gplfronttext{}%
  \makeatother
  \ifGPblacktext
    \def\colorrgb#1{}%
    \def\colorgray#1{}%
  \else
    \ifGPcolor
      \def\colorrgb#1{\color[rgb]{#1}}%
      \def\colorgray#1{\color[gray]{#1}}%
      \expandafter\def\csname LTw\endcsname{\color{white}}%
      \expandafter\def\csname LTb\endcsname{\color{black}}%
      \expandafter\def\csname LTa\endcsname{\color{black}}%
      \expandafter\def\csname LT0\endcsname{\color[rgb]{1,0,0}}%
      \expandafter\def\csname LT1\endcsname{\color[rgb]{0,1,0}}%
      \expandafter\def\csname LT2\endcsname{\color[rgb]{0,0,1}}%
      \expandafter\def\csname LT3\endcsname{\color[rgb]{1,0,1}}%
      \expandafter\def\csname LT4\endcsname{\color[rgb]{0,1,1}}%
      \expandafter\def\csname LT5\endcsname{\color[rgb]{1,1,0}}%
      \expandafter\def\csname LT6\endcsname{\color[rgb]{0,0,0}}%
      \expandafter\def\csname LT7\endcsname{\color[rgb]{1,0.3,0}}%
      \expandafter\def\csname LT8\endcsname{\color[rgb]{0.5,0.5,0.5}}%
    \else
      \def\colorrgb#1{\color{black}}%
      \def\colorgray#1{\color[gray]{#1}}%
      \expandafter\def\csname LTw\endcsname{\color{white}}%
      \expandafter\def\csname LTb\endcsname{\color{black}}%
      \expandafter\def\csname LTa\endcsname{\color{black}}%
      \expandafter\def\csname LT0\endcsname{\color{black}}%
      \expandafter\def\csname LT1\endcsname{\color{black}}%
      \expandafter\def\csname LT2\endcsname{\color{black}}%
      \expandafter\def\csname LT3\endcsname{\color{black}}%
      \expandafter\def\csname LT4\endcsname{\color{black}}%
      \expandafter\def\csname LT5\endcsname{\color{black}}%
      \expandafter\def\csname LT6\endcsname{\color{black}}%
      \expandafter\def\csname LT7\endcsname{\color{black}}%
      \expandafter\def\csname LT8\endcsname{\color{black}}%
    \fi
  \fi
  \setlength{\unitlength}{0.0500bp}%
  \begin{picture}(5096.00,7644.00)%
    \gplgaddtomacro\gplbacktext{%
      \csname LTb\endcsname%
      \put(487,414){\makebox(0,0)[r]{\strut{}$10^{48}$}}%
      \put(487,1446){\makebox(0,0)[r]{\strut{}$10^{49}$}}%
      \put(487,2479){\makebox(0,0)[r]{\strut{}$10^{50}$}}%
      \put(487,3511){\makebox(0,0)[r]{\strut{}$10^{51}$}}%
      \put(1442,252){\makebox(0,0){\strut{} 10}}%
      \put(2777,252){\makebox(0,0){\strut{} 100}}%
      \put(4111,252){\makebox(0,0){\strut{} 1000}}%
      \put(73,2118){\rotatebox{-270}{\makebox(0,0){\strut{}$\left|\Ebind\right|$ $(\rm erg)$}}}%
      \put(2776,72){\makebox(0,0){\strut{}Stellar Radius, $R_{\rm donor}$ $(\Rsun)$}}%
    }%
    \gplgaddtomacro\gplfronttext{%
    }%
    \gplgaddtomacro\gplbacktext{%
      \csname LTb\endcsname%
      \put(498,4401){\makebox(0,0)[r]{\strut{}0.01}}%
      \put(498,5509){\makebox(0,0)[r]{\strut{}0.1}}%
      \put(498,6616){\makebox(0,0)[r]{\strut{}1}}%
      \put(1442,7391){\makebox(0,0){\strut{} 10}}%
      \put(2777,7391){\makebox(0,0){\strut{} 100}}%
      \put(4111,7391){\makebox(0,0){\strut{} 1000}}%
      \put(84,5552){\rotatebox{-270}{\makebox(0,0){\strut{}$\lambda$}}}%
      \put(2776,3768){\makebox(0,0){\strut{}}}%
      \put(2776,7570){\makebox(0,0){\strut{}Stellar radius, $R_{\rm donor}$ $(\Rsun)$}}%
    }%
    \gplgaddtomacro\gplfronttext{%
      \csname LTb\endcsname%
      \put(1148,5047){\makebox(0,0)[l]{\strut{}MW}}%
      \csname LTb\endcsname%
      \put(1148,4867){\makebox(0,0)[l]{\strut{}$8\,\Msun$}}%
      \csname LTb\endcsname%
      \put(1148,4687){\makebox(0,0)[l]{\strut{}$15\,\Msun$}}%
      \csname LTb\endcsname%
      \put(1148,4507){\makebox(0,0)[l]{\strut{}$25\,\Msun$}}%
      \csname LTb\endcsname%
      \put(1148,4327){\makebox(0,0)[l]{\strut{}$40\,\Msun$}}%
      \csname LTb\endcsname%
      \put(1148,4147){\makebox(0,0)[l]{\strut{}$80\,\Msun$}}%
    }%
    \gplgaddtomacro\gplbacktext{%
      \csname LTb\endcsname%
      \put(2776,3768){\makebox(0,0){\strut{}}}%
    }%
    \gplgaddtomacro\gplfronttext{%
      \csname LTb\endcsname%
      \put(4413,6985){\makebox(0,0)[l]{\strut{}IZw18}}%
      \csname LTb\endcsname%
      \put(4413,6805){\makebox(0,0)[l]{\strut{}$8\,\Msun$}}%
      \csname LTb\endcsname%
      \put(4413,6625){\makebox(0,0)[l]{\strut{}$15\,\Msun$}}%
      \csname LTb\endcsname%
      \put(4413,6445){\makebox(0,0)[l]{\strut{}$26\,\Msun$}}%
      \csname LTb\endcsname%
      \put(4413,6265){\makebox(0,0)[l]{\strut{}$39\,\Msun$}}%
      \csname LTb\endcsname%
      \put(4413,6085){\makebox(0,0)[l]{\strut{}$88\,\Msun$}}%
    }%
    \gplbacktext
    \put(0,0){\includegraphics{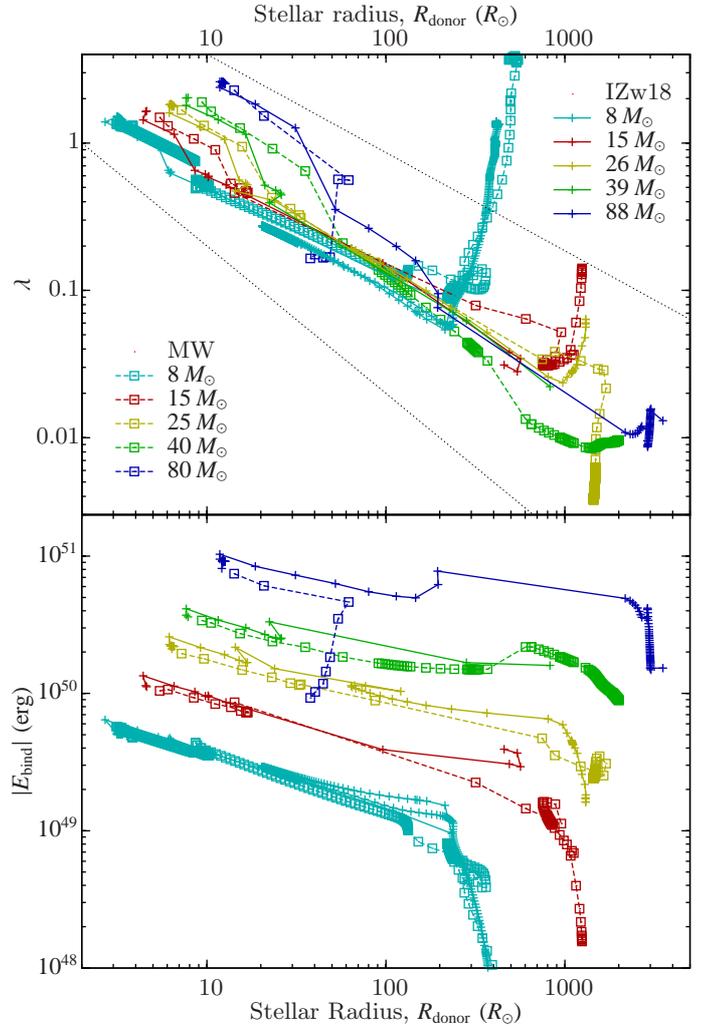}}%
    \gplfronttext
  \end{picture}%
\endgroup
\end{minipage}
  \caption[]{
     Binding energy of the envelope, $|E_{\rm bind}|$ (lower panel), and its associated $\lambda$-value
     (upper panel), as a function of total stellar radius for two sets of models with $Z=Z_{\odot}/50$ (full lines, crosses) and $Z=Z_{\rm MW}$ (dashed lines, squares).
     Independent of mass and metallicity, and before reaching the giant stages ($R\la1000\;R_{\odot}$), the $\lambda$-values almost follow a power law with an exponent 
     between $-2/3$ and $-1$ (upper and lower grey lines, respectively). 
     The exceptions are stars with initial masses $\ga 60\;M_{\odot}$ at $Z=Z_{\rm MW}$ (dashed blue line), which either become LBV stars, 
     see Sect.~\ref{subsec:LBV}, or have their envelopes stripped by enhanced wind-mass loss. 
     The absolute binding energies of the $8\;M_{\odot}$ stars drop below the plotted range, down to approximately $2.7\times 10^{47}\;{\rm erg}$ and $7.3\times 10^{46}\;{\rm erg}$ 
     for $Z=Z_{\odot}/50$ and $Z=Z_{\rm MW}$, respectively.
    }
\label{fig:Ebind}
\end{center}
\end{figure}

In Fig.~\ref{fig:Ebind} we have plotted our calculated values of $\lambda$ and $|E_{\rm bind}|$ according to Eq.~(\ref{eq:bindingenergy}) as a function of stellar radii, $R_{\rm donor}$,
using the stellar models of \citet{sly+15} for a metallicity of $Z=Z_{\odot}/50$ (resembling the metallicity of the irregular dwarf galaxy: I\,Zwicky~18) 
and \citet{bdc+11} for $Z=0.00876\approx Z_{\odot}/2$ (which we assume to represent the average metallicity of the Milky~Way, $Z_{\rm MW}$). 
The $8\;M_{\odot}$ models are calculated using the same code and input physics as the models of \citet{sly+15} and \citet{bdc+11}, respectively.

It is seen that, in general, the envelope becomes less bound with increasing values of $R_{\rm donor}$. 
Until the giant stages, the values of $|E_{\rm bind}|$ are moderately declining due to a combination of structural changes (i.e. growing core mass) and wind-mass loss,
which affects the mass-density profile and decreases the mass of the envelope. The evolution at these early stages is more or less independent of metallicity,
except for stars with ZAMS masses $\ga 60\;M_{\odot}$ , which become luminous blue variable (LBV) stars at high metallicities, cf. Sect.~\ref{subsec:LBV}, 
or have their envelopes stripped by enhanced wind mass loss \citep{vdl01}, and therefore do not evolve to become red supergiants 
(cf. the dashed blue track of the $80\;M_{\odot}$ star with $Z=Z_{\rm MW}$ , which does not expand above $60\;R_{\odot}$).  

The resulting change of $\lambda$ with stellar radius (upper panel) is seen to be significantly stronger than changes caused by different stellar masses or metallicities.
During the early stages of the expansion phase (up to $R\simeq 1000\;R_{\odot}$), the dependence on radius almost follows a power law with an exponent between $-2/3$ and $-1$.

The stellar tracks in Fig.~\ref{fig:Ebind} terminate at different evolutionary stages. Depending on stellar mass, the stars will reach the Eddington limit (Sect.~\ref{subsec:LBV}) at different epochs of evolution.
When a star reaches the Eddington limit, it initiates cycles of large-amplitude radial pulsations. This explains the horizontal clustering of points at the end of the stellar tracks
in the ($R,|E_{\rm bind}|)$--diagram. This also explains why some tracks have increasing $\lambda$--values near the end (e.g. $15\;M_{\odot}$ stars at $Z=Z_{\rm MW}$, evolved beyond core carbon burning), 
whereas others have decreasing values of $\lambda$ (e.g. $25\;M_{\odot}$ stars at $Z=Z_{\rm MW}$, only evolved to the end of core helium burning),
or more or less constant values of $\lambda$ (e.g. $40\;M_{\odot}$ stars at $Z=Z_{\rm MW}$, even less evolved to hydrogen shell burning). A careful inspection of the $15\;M_{\odot}$ track
at $Z=Z_{\rm MW}$ shows decreasing and increasing $\lambda$-values before reaching the pulsating stage as a giant.
This star experiences a final giant stage with significant radial expansion (up to $R=1585\;R_{\odot}$), which results in the strong decline in $|E_{\rm bind}|$, causing the increase in $\lambda$.

Our calculated $\lambda$-values for massive stars (with initial masses of up to $115\;M_{\odot}$) are in broad agreement with those
of \citet{dt01} and \citet{lvk11}. The latter authors demonstrated that the calculated $\lambda$-values are largely independent of 
metallicity and applied wind mass-loss prescription. Our calculations more or less confirm this result, except for massive metal-rich stars (LBVs)
or in case models are calculated with a {\it \textup{very}} low mass-loss rate (much lower than for $Z=Z_{\odot}/50$), in which case the $\lambda$-values become higher for evolved stars \citep{prh03}.

\begin{figure}[t]
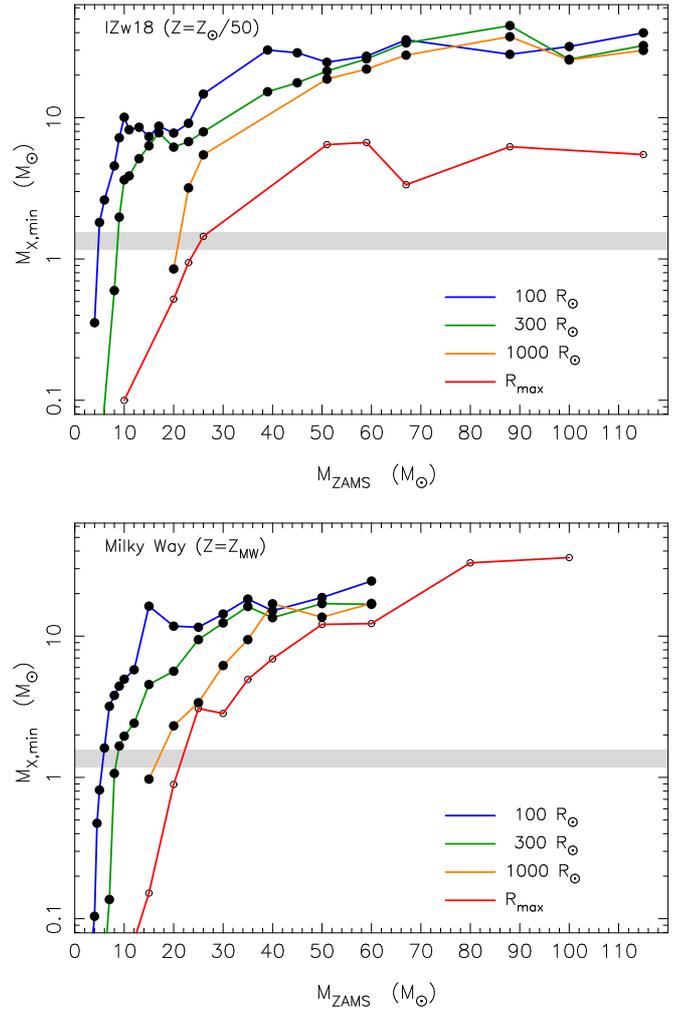

\begin{center}
\vspace{0.2cm}
 \mbox{\includegraphics[width=0.35\textwidth, angle=-90]{Mxmin2_IZw18.eps}}

\vspace{0.4cm}
 \mbox{\includegraphics[width=0.35\textwidth, angle=-90]{Mxmin2_MW.eps}}
  \caption[]{
    Minimum mass of the in-spiralling object, $M_{\rm X,min}$ which is needed to expel the envelope
    during a CE evolution with $\alpha=1$, for a given donor star radius as indicated by the coloured lines, 
    as a function of ZAMS mass of the donor star, $M_{\rm ZAMS}$
    for $Z=Z_{\odot}/50$ (upper panel) and $Z=Z_{\rm MW}$ (lower panel). $R_{\rm max}$ is the maximum radial extent during the stellar evolution.
    The grey band between 1.17 and $1.56\;M_{\odot}$ indicates masses of NSs observed in double NS systems \citep{msf+15}.
    }
\label{fig:Mxmin}
\end{center}
\end{figure}

\subsection{CE ejectability and companion star masses}\label{subsec:ejectability} 
To investigate the ejectability of the CE, we plot in Fig.~\ref{fig:Mxmin} the minimum mass of the in-spiralling object, $M_{\rm X,min}$
which is needed to successfully expel the envelope during a CE evolution of stars with a given mass and metallicity at
different evolutionary stages. 
The core radii of the stripped donor stars were calculated for naked helium star models (Sect.~\ref{subsec:He-env}) using the stellar 
evolution code BEC \citep{ywl10}, which was also used to calculate our applied models of \citet{sly+15} and \citet{bdc+11}. 

As expected from the decreasing values of $|E_{\rm bind}|$ with increasing stellar radius
(Fig.~\ref{fig:Ebind}), it is seen in Fig.~\ref{fig:Mxmin} that evolved (expanded) donor stars more easily have their
envelopes ejected by a relatively less massive in-spiralling companion than less evolved donor stars. In particular for the low-metallicity models
($Z=Z_{\odot}/50$, upper panel), we note the significant difference in envelope ejectability between massive stars evolved to radii  
$R\la 1000\;R_{\odot}$ (blue, green, and orange lines) and giants ($R>1000\;R_{\odot}$; red line).
For example, for the $88\;M_{\odot}$ model with $Z=Z_{\odot}/50$ we find that it requires an in-spiralling object with
a mass of at least $30-50\;M_{\odot}$ to eject the envelope in the former case, but only an object of $\sim\!6\;M_{\odot}$ 
in case the envelope is ejected when the donor star is an evolved giant near its maximum radius of $\sim\!3500\;R_{\odot}$.

The scatter of points along the coloured lines in Fig.~\ref{fig:Mxmin} can be understood from the non-monotonic
behaviour of $|E_{\rm bind}|$ as a function of stellar radius. For a given stellar mass, we see from
Fig.~\ref{fig:Ebind} that $|E_{\rm bind}|$ is not monotonically decreasing as a function of increasing value of $R$.
The reason for this are changes in the core structure during the stellar evolution.

We conclude that envelope ejection is facilitated for giant stars compared to less evolved stars, and as long as the in-spiralling BH masses 
are high enough (above the coloured lines in Fig.~\ref{fig:Mxmin}), they probably succeed in ejecting the envelopes of their
host stars. Hence, for a given donor star mass and mass of in-spiralling object, we can define a certain interval of stellar radii of each pre-CE donor star 
where CE ejection is possible (see Sect.~\ref{subsec:pop_syn}), which translates into a range of pre-CE orbital periods 
combining Kepler's~$ ^{\rm }$~third law with an expression for the dimensionless 
Roche-lobe radius of the donor star \citep[e.g.][]{egg83}. For example, our $115\;M_{\odot}$ model star ($Z=Z_{\odot}/50$) might have its envelope ejected successfully
when the mass of the in-spiralling BH is above $30-40\;M_{\odot}$ and the star has evolved to $R\ga100\;R_{\odot}$ (cf. Fig.~\ref{fig:Mxmin}, upper panel). 
The maximum radius reached by this star as a giant is $3922\;R_{\odot}$, at which point $M\simeq 93\;M_{\odot}$. Hence, the orbital period interval for successful
ejection of the envelope in this particular binary is between 875 and 7750~days for a $30\;M_{\odot}$ BH.

We stress the important caveat that the above calculations in Fig.~\ref{fig:Mxmin} all assume a certain core boundary criterion ($X_H=0.10$).
We investigate in Sect.~\ref{subsec:bifurcation} more carefully at which bifurcation point we might expect CE ejection. 
Furthermore, we assume formation of a CE for all binary systems in Fig.~\ref{fig:Mxmin}. It is quite likely that several of these
systems, especially with less evolved donor stars or mass ratios close to unity, may undergo stable Roche-lobe overflow \citep[RLO,][]{pibv16}. 
We also recall that in Fig.~\ref{fig:Mxmin} we solely investigate the possibility of CE ejection regardless of the formation of any given binary.
Some of the implied binary configurations in Fig.~\ref{fig:Mxmin} are unlikely to be produced in nature in an isolated binary system. For example,
it would be unexpected to have a $6\;M_{\odot}$ BH orbiting an $88\;M_{\odot}$ star (and, in particular, a 100 or a $115\;M_{\odot}$ star).
The reason is that in order for the primary star (the initially most massive of the two ZAMS stars) to evolve first and eventually produce a BH, it
must have had a ZAMS mass of more than $\sim\!60\;M_{\odot}$ (otherwise the initially least massive of the two stars, the secondary,
would not be able to accrete sufficient material to reach $88\;M_{\odot}$). However, the core mass of a $60\;M_{\odot}$ star is $\sim\!30\;M_{\odot}$ 
and thus most likely too massive to leave a BH with a mass of only $6\;M_{\odot}$.

Finally, we note that the cores of the most massive ($\ge 100\;M_{\odot}$) low-metallicity stars exceed $60\;M_{\odot}$ and probably terminate their
lives in pair-instability SNe, which lead to the total disruption of the star without leaving behind any BH \citep{hw02,cw12}.

All calculations in Fig.~\ref{fig:Mxmin} were performed assuming an envelope ejection efficiency parameter
of $\alpha=1$. The plotted curves therefore represent the most optimistic case for ejectability. Applying more realistic efficiencies of $\alpha<1$
would require higher values of $M_{\rm X,min}$ and shift all plotted curves upward. 
On the other hand, we assumed that the release of orbital binding energy is the sole energy source available to eject the envelope.
It is possible that other energy sources are at work as well (see Sect.~\ref{subsec:other_sources}), in which case the curves in Fig.~\ref{fig:Mxmin} would be
shifted downward to lower values of $M_{\rm X,min}$, reflecting that envelope ejection would be facilitated. 

\subsection{CE ejection: NS-NS and WD-WD binaries}\label{subsec:NSNSWDWD} 
Another interesting result seen in Fig.~\ref{fig:Mxmin} is that HMXB systems with in-spiralling NSs 
are also able to eject the envelopes of donor stars with initial masses of up to about $22-25\;M_{\odot}$ (depending on metallicity).
These systems eventually evolve to become double NS systems following a post-CE episode of so-called Case~BB Roche-lobe overflow \citep{dpsv02,ibk+03,tlp15}.
The grey band between 1.17 and $1.56\;M_{\odot}$ in Fig.~\ref{fig:Mxmin} indicates the interval of measured NS masses in double NS systems \citep{msf+15}.
Similarly, we note that evolved donors with masses lower than $8-10\;M_{\odot}$ can have their envelopes ejected by even sub-solar mass objects,
thereby allowing formation of tight double WD systems through CE evolution, as confirmed by observations \citep[e.g.][]{zsgn10}. 

Interestingly enough, \citet{ijp15} found that less massive in-spiralling stars plunge in faster than more massive in-spiralling stars, which results in a relatively higher
heating rate, and low-mass intruders are therefore more effective in ejecting the envelopes since a smaller fraction of the released 
orbital energy is dissipated in the outer parts of the envelope. This suggests that low-mass stars are more efficient in removing the
envelope \citep[i.e. these binaries should have higher $\alpha$-values than binaries with more massive companion stars, see also][]{pod01}.
This hypothesis is supported by observations that indicate that the ejection efficiency increases for less massive companion stars \citep{dpm+11,dkk12}.

\vspace{0.5cm}
\subsection{Bifurcation point of envelope ejection}\label{subsec:bifurcation}
One of the main problems in our understanding of CE ejection is the difficulty of localising the physical point of envelope ejection, that is,
the bifurcation point, which separates the ejected envelope from the remaining core \citep{td01}.
The three main categories proposed for determining the bifurcation point are nuclear energy generation, chemical composition, and thermodynamic quantities.
A first-order constraint on the location of the core boundary (i.e. bifurcation point) can be taken as follows:
it has to be somewhere between the hydrogen-depleted core ($X_H=0$) and the mass coordinate of the bottom of the convection zone in the pre-CE star.
From studies of direct collisions between a NS and a red giant, it was found \citep{lpd+06} that some amount of hydrogen
remains bound to the stellar core, following envelope ejection. This result therefore indicates $X_H>0$.
The problem with the upper limit is that the bottom of the outer convection zone often moves in mass coordinate during the CE ejection.

Our chosen core boundary criterion ($X_H=0.10$) is easy to apply in practice to stars at most evolutionary stages.
Changing the mass coordinate of the core-mass boundary by 1\% results in only minor different values of $E_{\rm bind}$ and $\lambda$ (of the order 10\%)
for stars in the Hertzsprung gap, whereas the effect of changing $M_{\rm core}$ by 1\% is much larger (up to a factor 2) for stars on the giant branch 
that possess a steep density gradient near the core boundary.

\begin{figure}[t]
\begin{center}
\begin{minipage}{\columnwidth}
\begingroup
  \fontfamily{Hevetica}%
  \selectfont
  \makeatletter
  \providecommand\color[2][]{%
    \GenericError{(gnuplot) \space\space\space\@spaces}{%
      Package color not loaded in conjunction with
      terminal option `colourtext'%
    }{See the gnuplot documentation for explanation.%
    }{Either use 'blacktext' in gnuplot or load the package
      color.sty in LaTeX.}%
    \renewcommand\color[2][]{}%
  }%
  \providecommand\includegraphics[2][]{%
    \GenericError{(gnuplot) \space\space\space\@spaces}{%
      Package graphicx or graphics not loaded%
    }{See the gnuplot documentation for explanation.%
    }{The gnuplot epslatex terminal needs graphicx.sty or graphics.sty.}%
    \renewcommand\includegraphics[2][]{}%
  }%
  \providecommand\rotatebox[2]{#2}%
  \@ifundefined{ifGPcolor}{%
    \newif\ifGPcolor
    \GPcolortrue
  }{}%
  \@ifundefined{ifGPblacktext}{%
    \newif\ifGPblacktext
    \GPblacktexttrue
  }{}%
  \let\gplgaddtomacro\g@addto@macro
  \gdef\gplbacktext{}%
  \gdef\gplfronttext{}%
  \makeatother
  \ifGPblacktext
    \def\colorrgb#1{}%
    \def\colorgray#1{}%
  \else
    \ifGPcolor
      \def\colorrgb#1{\color[rgb]{#1}}%
      \def\colorgray#1{\color[gray]{#1}}%
      \expandafter\def\csname LTw\endcsname{\color{white}}%
      \expandafter\def\csname LTb\endcsname{\color{black}}%
      \expandafter\def\csname LTa\endcsname{\color{black}}%
      \expandafter\def\csname LT0\endcsname{\color[rgb]{1,0,0}}%
      \expandafter\def\csname LT1\endcsname{\color[rgb]{0,1,0}}%
      \expandafter\def\csname LT2\endcsname{\color[rgb]{0,0,1}}%
      \expandafter\def\csname LT3\endcsname{\color[rgb]{1,0,1}}%
      \expandafter\def\csname LT4\endcsname{\color[rgb]{0,1,1}}%
      \expandafter\def\csname LT5\endcsname{\color[rgb]{1,1,0}}%
      \expandafter\def\csname LT6\endcsname{\color[rgb]{0,0,0}}%
      \expandafter\def\csname LT7\endcsname{\color[rgb]{1,0.3,0}}%
      \expandafter\def\csname LT8\endcsname{\color[rgb]{0.5,0.5,0.5}}%
    \else
      \def\colorrgb#1{\color{black}}%
      \def\colorgray#1{\color[gray]{#1}}%
      \expandafter\def\csname LTw\endcsname{\color{white}}%
      \expandafter\def\csname LTb\endcsname{\color{black}}%
      \expandafter\def\csname LTa\endcsname{\color{black}}%
      \expandafter\def\csname LT0\endcsname{\color{black}}%
      \expandafter\def\csname LT1\endcsname{\color{black}}%
      \expandafter\def\csname LT2\endcsname{\color{black}}%
      \expandafter\def\csname LT3\endcsname{\color{black}}%
      \expandafter\def\csname LT4\endcsname{\color{black}}%
      \expandafter\def\csname LT5\endcsname{\color{black}}%
      \expandafter\def\csname LT6\endcsname{\color{black}}%
      \expandafter\def\csname LT7\endcsname{\color{black}}%
      \expandafter\def\csname LT8\endcsname{\color{black}}%
    \fi
  \fi
  \setlength{\unitlength}{0.0500bp}%
  \begin{picture}(5096.00,3822.00)%
    \gplgaddtomacro\gplbacktext{%
      \csname LTb\endcsname%
      \put(487,909){\makebox(0,0)[r]{\strut{}$10^{48}$}}%
      \put(487,1565){\makebox(0,0)[r]{\strut{}$10^{49}$}}%
      \put(487,2221){\makebox(0,0)[r]{\strut{}$10^{50}$}}%
      \put(487,2877){\makebox(0,0)[r]{\strut{}$10^{51}$}}%
      \put(487,3533){\makebox(0,0)[r]{\strut{}$10^{52}$}}%
      \put(509,288){\makebox(0,0){\strut{}$0$}}%
      \put(1002,288){\makebox(0,0){\strut{}$10$}}%
      \put(1495,288){\makebox(0,0){\strut{}$20$}}%
      \put(1988,288){\makebox(0,0){\strut{}$30$}}%
      \put(2481,288){\makebox(0,0){\strut{}$40$}}%
      \put(2974,288){\makebox(0,0){\strut{}$50$}}%
      \put(3467,288){\makebox(0,0){\strut{}$60$}}%
      \put(3960,288){\makebox(0,0){\strut{}$70$}}%
      \put(4452,288){\makebox(0,0){\strut{}$80$}}%
      \put(4945,288){\makebox(0,0){\strut{}$90$}}%
      \put(73,2090){\rotatebox{-270}{\makebox(0,0){\strut{}$-E~(\rm erg)$}}}%
      \put(2776,108){\makebox(0,0){\strut{}$\text{Mass coordinate,}~m~(\Msun)$}}%
    }%
    \gplgaddtomacro\gplfronttext{%
      \csname LTb\endcsname%
      \put(4549,3432){\makebox(0,0)[r]{\strut{}convection}}%
      \csname LTb\endcsname%
      \put(4549,3162){\makebox(0,0)[r]{\strut{}$E_{\rm grav}$}}%
      \csname LTb\endcsname%
      \put(4549,2892){\makebox(0,0)[r]{\strut{}$E_{\rm bind}$}}%
    }%
    \gplgaddtomacro\gplbacktext{%
      \csname LTb\endcsname%
      \put(487,909){\makebox(0,0)[r]{\strut{}\phantom{$10^{48}$}}}%
      \put(487,1565){\makebox(0,0)[r]{\strut{}\phantom{$10^{49}$}}}%
      \put(487,2221){\makebox(0,0)[r]{\strut{}\phantom{$10^{50}$}}}%
      \put(487,2877){\makebox(0,0)[r]{\strut{}\phantom{$10^{51}$}}}%
      \put(487,3533){\makebox(0,0)[r]{\strut{}\phantom{$10^{52}$}}}%
      \put(509,288){\makebox(0,0){\strut{}\phantom{$0$}}}%
      \put(1002,288){\makebox(0,0){\strut{}\phantom{$10$}}}%
      \put(1495,288){\makebox(0,0){\strut{}\phantom{$20$}}}%
      \put(1988,288){\makebox(0,0){\strut{}\phantom{$30$}}}%
      \put(2481,288){\makebox(0,0){\strut{}\phantom{$40$}}}%
      \put(2974,288){\makebox(0,0){\strut{}\phantom{$50$}}}%
      \put(3467,288){\makebox(0,0){\strut{}\phantom{$60$}}}%
      \put(3960,288){\makebox(0,0){\strut{}\phantom{$70$}}}%
      \put(4452,288){\makebox(0,0){\strut{}\phantom{$80$}}}%
      \put(4945,288){\makebox(0,0){\strut{}\phantom{$90$}}}%
      \put(-35,2090){\rotatebox{-270}{\makebox(0,0){\strut{}\phantom{$-E~(\rm erg)$}}}}%
      \put(2776,108){\makebox(0,0){\strut{}\phantom{$\text{Mass coordinate,}~m~(\Msun)$}}}%
    }%
    \gplgaddtomacro\gplfronttext{%
      \csname LTb\endcsname%
      \put(1208,1398){\makebox(0,0)[l]{\strut{}$\Delta E_{{\rm orb,}1.3}$}}%
      \csname LTb\endcsname%
      \put(1208,1128){\makebox(0,0)[l]{\strut{}$\Delta E_{{\rm orb,}5}$}}%
      \csname LTb\endcsname%
      \put(1208,858){\makebox(0,0)[l]{\strut{}$\Delta E_{{\rm orb,}10}$}}%
    }%
    \gplgaddtomacro\gplbacktext{%
      \csname LTb\endcsname%
      \put(487,909){\makebox(0,0)[r]{\strut{}\phantom{$10^{48}$}}}%
      \put(487,1565){\makebox(0,0)[r]{\strut{}\phantom{$10^{49}$}}}%
      \put(487,2221){\makebox(0,0)[r]{\strut{}\phantom{$10^{50}$}}}%
      \put(487,2877){\makebox(0,0)[r]{\strut{}\phantom{$10^{51}$}}}%
      \put(487,3533){\makebox(0,0)[r]{\strut{}\phantom{$10^{52}$}}}%
      \put(509,288){\makebox(0,0){\strut{}\phantom{$0$}}}%
      \put(1002,288){\makebox(0,0){\strut{}\phantom{$10$}}}%
      \put(1495,288){\makebox(0,0){\strut{}\phantom{$20$}}}%
      \put(1988,288){\makebox(0,0){\strut{}\phantom{$30$}}}%
      \put(2481,288){\makebox(0,0){\strut{}\phantom{$40$}}}%
      \put(2974,288){\makebox(0,0){\strut{}\phantom{$50$}}}%
      \put(3467,288){\makebox(0,0){\strut{}\phantom{$60$}}}%
      \put(3960,288){\makebox(0,0){\strut{}\phantom{$70$}}}%
      \put(4452,288){\makebox(0,0){\strut{}\phantom{$80$}}}%
      \put(4945,288){\makebox(0,0){\strut{}\phantom{$90$}}}%
      \put(-35,2090){\rotatebox{-270}{\makebox(0,0){\strut{}\phantom{$-E~(\rm erg)$}}}}%
      \put(2776,108){\makebox(0,0){\strut{}\phantom{$\text{Mass coordinate,}~m~(\Msun)$}}}%
    }%
    \gplgaddtomacro\gplfronttext{%
      \csname LTb\endcsname%
      \put(2405,1398){\makebox(0,0)[l]{\strut{}$\Delta E_{{\rm orb,}20}$}}%
      \csname LTb\endcsname%
      \put(2405,1128){\makebox(0,0)[l]{\strut{}$\Delta E_{{\rm orb,}40}$}}%
      \csname LTb\endcsname%
      \put(2405,858){\makebox(0,0)[l]{\strut{}$\Delta E_{{\rm orb,}80}$}}%
    }%
    \gplgaddtomacro\gplbacktext{%
      \csname LTb\endcsname%
      \put(487,909){\makebox(0,0)[r]{\strut{}\phantom{$10^{48}$}}}%
      \put(487,1565){\makebox(0,0)[r]{\strut{}\phantom{$10^{49}$}}}%
      \put(487,2221){\makebox(0,0)[r]{\strut{}\phantom{$10^{50}$}}}%
      \put(487,2877){\makebox(0,0)[r]{\strut{}\phantom{$10^{51}$}}}%
      \put(487,3533){\makebox(0,0)[r]{\strut{}\phantom{$10^{52}$}}}%
      \put(509,288){\makebox(0,0){\strut{}\phantom{$0$}}}%
      \put(1002,288){\makebox(0,0){\strut{}\phantom{$10$}}}%
      \put(1495,288){\makebox(0,0){\strut{}\phantom{$20$}}}%
      \put(1988,288){\makebox(0,0){\strut{}\phantom{$30$}}}%
      \put(2481,288){\makebox(0,0){\strut{}\phantom{$40$}}}%
      \put(2974,288){\makebox(0,0){\strut{}\phantom{$50$}}}%
      \put(3467,288){\makebox(0,0){\strut{}\phantom{$60$}}}%
      \put(3960,288){\makebox(0,0){\strut{}\phantom{$70$}}}%
      \put(4452,288){\makebox(0,0){\strut{}\phantom{$80$}}}%
      \put(4945,288){\makebox(0,0){\strut{}\phantom{$90$}}}%
      \put(-35,2090){\rotatebox{-270}{\makebox(0,0){\strut{}\phantom{$-E~(\rm erg)$}}}}%
      \put(2776,108){\makebox(0,0){\strut{}\phantom{$\text{Mass coordinate,}~m~(\Msun)$}}}%
    }%
    \gplgaddtomacro\gplfronttext{%
    }%
    \gplbacktext
    \put(0,0){\includegraphics{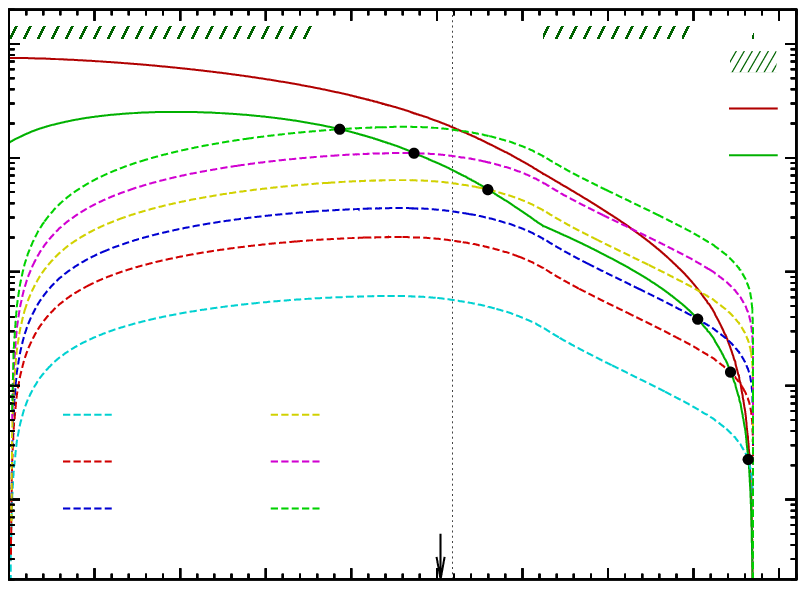}}%
    \gplfronttext
  \end{picture}%
\endgroup

\begingroup
  \fontfamily{Hevetica}%
  \selectfont
  \makeatletter
  \providecommand\color[2][]{%
    \GenericError{(gnuplot) \space\space\space\@spaces}{%
      Package color not loaded in conjunction with
      terminal option `colourtext'%
    }{See the gnuplot documentation for explanation.%
    }{Either use 'blacktext' in gnuplot or load the package
      color.sty in LaTeX.}%
    \renewcommand\color[2][]{}%
  }%
  \providecommand\includegraphics[2][]{%
    \GenericError{(gnuplot) \space\space\space\@spaces}{%
      Package graphicx or graphics not loaded%
    }{See the gnuplot documentation for explanation.%
    }{The gnuplot epslatex terminal needs graphicx.sty or graphics.sty.}%
    \renewcommand\includegraphics[2][]{}%
  }%
  \providecommand\rotatebox[2]{#2}%
  \@ifundefined{ifGPcolor}{%
    \newif\ifGPcolor
    \GPcolortrue
  }{}%
  \@ifundefined{ifGPblacktext}{%
    \newif\ifGPblacktext
    \GPblacktexttrue
  }{}%
  \let\gplgaddtomacro\g@addto@macro
  \gdef\gplbacktext{}%
  \gdef\gplfronttext{}%
  \makeatother
  \ifGPblacktext
    \def\colorrgb#1{}%
    \def\colorgray#1{}%
  \else
    \ifGPcolor
      \def\colorrgb#1{\color[rgb]{#1}}%
      \def\colorgray#1{\color[gray]{#1}}%
      \expandafter\def\csname LTw\endcsname{\color{white}}%
      \expandafter\def\csname LTb\endcsname{\color{black}}%
      \expandafter\def\csname LTa\endcsname{\color{black}}%
      \expandafter\def\csname LT0\endcsname{\color[rgb]{1,0,0}}%
      \expandafter\def\csname LT1\endcsname{\color[rgb]{0,1,0}}%
      \expandafter\def\csname LT2\endcsname{\color[rgb]{0,0,1}}%
      \expandafter\def\csname LT3\endcsname{\color[rgb]{1,0,1}}%
      \expandafter\def\csname LT4\endcsname{\color[rgb]{0,1,1}}%
      \expandafter\def\csname LT5\endcsname{\color[rgb]{1,1,0}}%
      \expandafter\def\csname LT6\endcsname{\color[rgb]{0,0,0}}%
      \expandafter\def\csname LT7\endcsname{\color[rgb]{1,0.3,0}}%
      \expandafter\def\csname LT8\endcsname{\color[rgb]{0.5,0.5,0.5}}%
    \else
      \def\colorrgb#1{\color{black}}%
      \def\colorgray#1{\color[gray]{#1}}%
      \expandafter\def\csname LTw\endcsname{\color{white}}%
      \expandafter\def\csname LTb\endcsname{\color{black}}%
      \expandafter\def\csname LTa\endcsname{\color{black}}%
      \expandafter\def\csname LT0\endcsname{\color{black}}%
      \expandafter\def\csname LT1\endcsname{\color{black}}%
      \expandafter\def\csname LT2\endcsname{\color{black}}%
      \expandafter\def\csname LT3\endcsname{\color{black}}%
      \expandafter\def\csname LT4\endcsname{\color{black}}%
      \expandafter\def\csname LT5\endcsname{\color{black}}%
      \expandafter\def\csname LT6\endcsname{\color{black}}%
      \expandafter\def\csname LT7\endcsname{\color{black}}%
      \expandafter\def\csname LT8\endcsname{\color{black}}%
    \fi
  \fi
  \setlength{\unitlength}{0.0500bp}%
  \begin{picture}(5096.00,3822.00)%
    \gplgaddtomacro\gplbacktext{%
      \csname LTb\endcsname%
      \put(487,750){\makebox(0,0)[r]{\strut{}$10^{45}$}}%
      \put(487,1178){\makebox(0,0)[r]{\strut{}$10^{46}$}}%
      \put(487,1607){\makebox(0,0)[r]{\strut{}$10^{47}$}}%
      \put(487,2036){\makebox(0,0)[r]{\strut{}$10^{48}$}}%
      \put(487,2464){\makebox(0,0)[r]{\strut{}$10^{49}$}}%
      \put(487,2893){\makebox(0,0)[r]{\strut{}$10^{50}$}}%
      \put(487,3322){\makebox(0,0)[r]{\strut{}$10^{51}$}}%
      \put(600,288){\makebox(0,0){\strut{}$30$}}%
      \put(1053,288){\makebox(0,0){\strut{}$35$}}%
      \put(1507,288){\makebox(0,0){\strut{}$40$}}%
      \put(1960,288){\makebox(0,0){\strut{}$45$}}%
      \put(2414,288){\makebox(0,0){\strut{}$50$}}%
      \put(2867,288){\makebox(0,0){\strut{}$55$}}%
      \put(3321,288){\makebox(0,0){\strut{}$60$}}%
      \put(3774,288){\makebox(0,0){\strut{}$65$}}%
      \put(4228,288){\makebox(0,0){\strut{}$70$}}%
      \put(4681,288){\makebox(0,0){\strut{}$75$}}%
      \put(73,2090){\rotatebox{-270}{\makebox(0,0){\strut{}$-E~(\rm erg)$}}}%
      \put(2776,108){\makebox(0,0){\strut{}$\text{Mass coordinate,}~m~(\Msun)$}}%
    }%
    \gplgaddtomacro\gplfronttext{%
      \csname LTb\endcsname%
      \put(4549,3432){\makebox(0,0)[r]{\strut{}convection}}%
      \csname LTb\endcsname%
      \put(4549,3162){\makebox(0,0)[r]{\strut{}$E_{\rm grav}$}}%
      \csname LTb\endcsname%
      \put(4549,2892){\makebox(0,0)[r]{\strut{}$E_{\rm bind}$}}%
    }%
    \gplgaddtomacro\gplbacktext{%
      \csname LTb\endcsname%
      \put(487,750){\makebox(0,0)[r]{\strut{}\phantom{$10^{45}$}}}%
      \put(487,1178){\makebox(0,0)[r]{\strut{}\phantom{$10^{46}$}}}%
      \put(487,1607){\makebox(0,0)[r]{\strut{}\phantom{$10^{47}$}}}%
      \put(487,2036){\makebox(0,0)[r]{\strut{}\phantom{$10^{48}$}}}%
      \put(487,2464){\makebox(0,0)[r]{\strut{}\phantom{$10^{49}$}}}%
      \put(487,2893){\makebox(0,0)[r]{\strut{}\phantom{$10^{50}$}}}%
      \put(487,3322){\makebox(0,0)[r]{\strut{}\phantom{$10^{51}$}}}%
      \put(600,288){\makebox(0,0){\strut{}\phantom{$30$}}}%
      \put(1053,288){\makebox(0,0){\strut{}\phantom{$35$}}}%
      \put(1507,288){\makebox(0,0){\strut{}\phantom{$40$}}}%
      \put(1960,288){\makebox(0,0){\strut{}\phantom{$45$}}}%
      \put(2414,288){\makebox(0,0){\strut{}\phantom{$50$}}}%
      \put(2867,288){\makebox(0,0){\strut{}\phantom{$55$}}}%
      \put(3321,288){\makebox(0,0){\strut{}\phantom{$60$}}}%
      \put(3774,288){\makebox(0,0){\strut{}\phantom{$65$}}}%
      \put(4228,288){\makebox(0,0){\strut{}\phantom{$70$}}}%
      \put(4681,288){\makebox(0,0){\strut{}\phantom{$75$}}}%
      \put(-35,2090){\rotatebox{-270}{\makebox(0,0){\strut{}\phantom{$-E~(\rm erg)$}}}}%
      \put(2776,108){\makebox(0,0){\strut{}\phantom{$\text{Mass coordinate,}~m~(\Msun)$}}}%
    }%
    \gplgaddtomacro\gplfronttext{%
      \csname LTb\endcsname%
      \put(1208,2218){\makebox(0,0)[l]{\strut{}$\Delta E_{{\rm orb,}1.3}$}}%
      \csname LTb\endcsname%
      \put(1208,1948){\makebox(0,0)[l]{\strut{}$\Delta E_{{\rm orb,}5}$}}%
      \csname LTb\endcsname%
      \put(1208,1678){\makebox(0,0)[l]{\strut{}$\Delta E_{{\rm orb,}10}$}}%
      \csname LTb\endcsname%
      \put(1208,1408){\makebox(0,0)[l]{\strut{}$\Delta E_{{\rm orb,}20}$}}%
      \csname LTb\endcsname%
      \put(1208,1138){\makebox(0,0)[l]{\strut{}$\Delta E_{{\rm orb,}40}$}}%
      \csname LTb\endcsname%
      \put(1208,868){\makebox(0,0)[l]{\strut{}$\Delta E_{{\rm orb,}80}$}}%
    }%
    \gplgaddtomacro\gplbacktext{%
      \csname LTb\endcsname%
      \put(487,750){\makebox(0,0)[r]{\strut{}\phantom{$10^{45}$}}}%
      \put(487,1178){\makebox(0,0)[r]{\strut{}\phantom{$10^{46}$}}}%
      \put(487,1607){\makebox(0,0)[r]{\strut{}\phantom{$10^{47}$}}}%
      \put(487,2036){\makebox(0,0)[r]{\strut{}\phantom{$10^{48}$}}}%
      \put(487,2464){\makebox(0,0)[r]{\strut{}\phantom{$10^{49}$}}}%
      \put(487,2893){\makebox(0,0)[r]{\strut{}\phantom{$10^{50}$}}}%
      \put(487,3322){\makebox(0,0)[r]{\strut{}\phantom{$10^{51}$}}}%
      \put(600,288){\makebox(0,0){\strut{}\phantom{$30$}}}%
      \put(1053,288){\makebox(0,0){\strut{}\phantom{$35$}}}%
      \put(1507,288){\makebox(0,0){\strut{}\phantom{$40$}}}%
      \put(1960,288){\makebox(0,0){\strut{}\phantom{$45$}}}%
      \put(2414,288){\makebox(0,0){\strut{}\phantom{$50$}}}%
      \put(2867,288){\makebox(0,0){\strut{}\phantom{$55$}}}%
      \put(3321,288){\makebox(0,0){\strut{}\phantom{$60$}}}%
      \put(3774,288){\makebox(0,0){\strut{}\phantom{$65$}}}%
      \put(4228,288){\makebox(0,0){\strut{}\phantom{$70$}}}%
      \put(4681,288){\makebox(0,0){\strut{}\phantom{$75$}}}%
      \put(-35,2090){\rotatebox{-270}{\makebox(0,0){\strut{}\phantom{$-E~(\rm erg)$}}}}%
      \put(2776,108){\makebox(0,0){\strut{}\phantom{$\text{Mass coordinate,}~m~(\Msun)$}}}%
    }%
    \gplgaddtomacro\gplfronttext{%
    }%
    \gplgaddtomacro\gplbacktext{%
      \csname LTb\endcsname%
      \put(487,750){\makebox(0,0)[r]{\strut{}\phantom{$10^{45}$}}}%
      \put(487,1178){\makebox(0,0)[r]{\strut{}\phantom{$10^{46}$}}}%
      \put(487,1607){\makebox(0,0)[r]{\strut{}\phantom{$10^{47}$}}}%
      \put(487,2036){\makebox(0,0)[r]{\strut{}\phantom{$10^{48}$}}}%
      \put(487,2464){\makebox(0,0)[r]{\strut{}\phantom{$10^{49}$}}}%
      \put(487,2893){\makebox(0,0)[r]{\strut{}\phantom{$10^{50}$}}}%
      \put(487,3322){\makebox(0,0)[r]{\strut{}\phantom{$10^{51}$}}}%
      \put(600,288){\makebox(0,0){\strut{}\phantom{$30$}}}%
      \put(1053,288){\makebox(0,0){\strut{}\phantom{$35$}}}%
      \put(1507,288){\makebox(0,0){\strut{}\phantom{$40$}}}%
      \put(1960,288){\makebox(0,0){\strut{}\phantom{$45$}}}%
      \put(2414,288){\makebox(0,0){\strut{}\phantom{$50$}}}%
      \put(2867,288){\makebox(0,0){\strut{}\phantom{$55$}}}%
      \put(3321,288){\makebox(0,0){\strut{}\phantom{$60$}}}%
      \put(3774,288){\makebox(0,0){\strut{}\phantom{$65$}}}%
      \put(4228,288){\makebox(0,0){\strut{}\phantom{$70$}}}%
      \put(4681,288){\makebox(0,0){\strut{}\phantom{$75$}}}%
      \put(-35,2090){\rotatebox{-270}{\makebox(0,0){\strut{}\phantom{$-E~(\rm erg)$}}}}%
      \put(2776,108){\makebox(0,0){\strut{}\phantom{$\text{Mass coordinate,}~m~(\Msun)$}}}%
    }%
    \gplgaddtomacro\gplfronttext{%
    }%
    \gplbacktext
    \put(0,0){\includegraphics{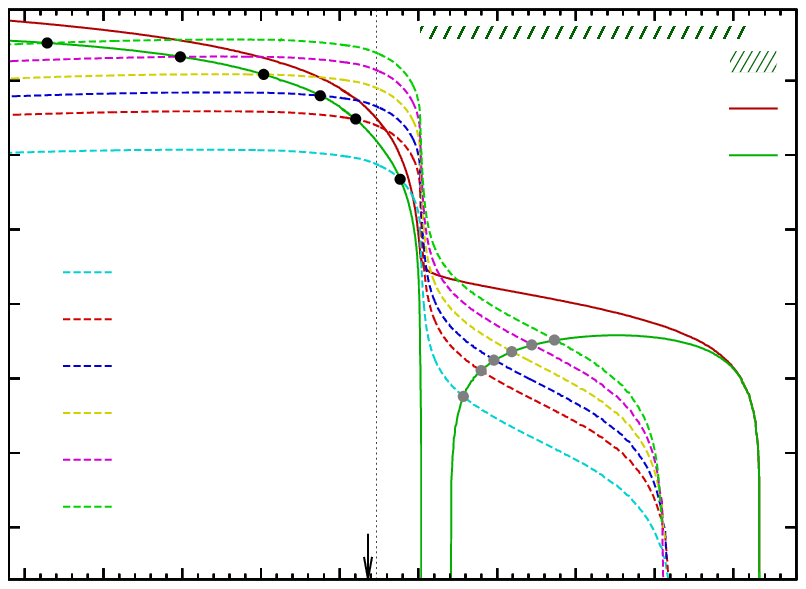}}%
    \gplfronttext
  \end{picture}%
\endgroup
\end{minipage}
  \caption[]{
Energy budget of the in-spiral process for the $88\;M_{\odot}$ donor star model with $Z=Z_{\odot}/50$ at two different evolutionary epochs:
at the beginning of the Hertzsprung gap, $R=194\;R_{\odot}$ (upper panel), and when the star reaches its maximum extent as a giant star, $R=3530\;R_{\odot}$ (lower panel).
The two solid lines mark the integrated binding energy of the material between the given mass coordinate and the surface. 
The green line ($E_{\rm bind}$) includes the total energy (internal and gravitational binding energy, cf. Eq.~(\ref{eq:bindingenergy})), while 
the red line ($E_{\rm grav}$) only considers the gravitational binding energy. 
The six dashed curves represent the released orbital energy, $\Delta E_{\rm orb}$ of an in-spiralling object with a given mass between 1.3 and $80\;M_{\odot}$ (cf. mass values in the legend). 
For the values of $\Delta E_{\rm orb}$, it is assumed that the in-spiral stops just when the remaining star would fill its Roche~lobe.
The vertical dotted line indicates the core boundary according to the $X_H=0.10$ criterion (cf. Sect.~\ref{sec:CE_ejection}), and the arrow marks the location of
the maximum-compression point (cf. Sect.~\ref{subsec:bif_revisited}). 
The hatched regions shown at the top indicate convective layers.
The black and grey dots mark the crossings when $E_{\rm bind}=\Delta E_{\rm orb}$, see Sect.~\ref{subsec:bifurcation} for details. 
    }
\label{fig:structure}
\end{center}
\end{figure}

In Fig.~\ref{fig:structure} we plot the integrated binding energy (solid lines) and the released orbital energy (dashed lines, calculated as the difference between $E_{\rm orb}$ at the
starting point and the end point of the in-spiral)
as a function of mass coordinate of our $88\;M_{\odot}$ stellar model ($Z=Z_{\odot}/50$) at two different evolutionary epochs of the star.
The in-spiralling object corresponds here to either a NS (with a mass of $1.3\;M_{\odot}$) or a BH with a mass between 5 and $80\;M_{\odot}$.
The upper panel is based on the structure of the star for $R=194.5\;R_{\odot}$ ($M=86.94\;M_{\odot}$, $M_{\rm core}=51.82\;M_{\odot}$)
during hydrogen shell burning (Hertzsprung gap star), while the lower panel is for $R=3530\;R_{\odot}$ ($M=76.65\;M_{\odot}$, $M_{\rm core}=52.35\;M_{\odot}$)
at its maximum expansion point as a giant.

In each panel in Fig.~\ref{fig:structure}, the black points indicate when (during in-spiral) the released orbital energy will become lower than the binding energy
of the layers outside the location of these points. Hence, if the in-spiralling object moves further inward than these crossing points, there is
no possibility of ejecting the envelope, and the system merges. The location of these black crossing points should be compared to our defined
location of the core boundary, shown as a vertical dotted line ($X_H=0.10$). If the black crossing points are located outside (to the right of) our assumed core boundary,
the system is expected to merge. In the upper panel (Hertzsprung gap star), this is the case for $M_{\rm X}=1.3-20\;M_{\odot}$, whereas $M_{\rm X}=40\;M_{\odot}$
and $80\;M_{\odot}$ succeed in envelope ejection. In the lower panel (star at the tip of the giant branch), all in-spiralling objects with masses $M_{\rm X}\ga5\;M_{\odot}$
are in principle able to eject the envelope of the star\footnote{In Fig.~\ref{fig:Mxmin} the limiting mass for the same model is slightly higher (about $6\;M_{\odot}$)
because of differences in estimating the core radius. In Fig.~\ref{fig:Mxmin} we applied naked (post-CE) helium star models (i.e. with zero pressure at their surfaces), whereas
Fig.~\ref{fig:structure}  probes the interior structure before envelope removal (i.e. with a non-zero surface pressure at a given point from the surrounding outer layers).
Hence, in the latter case the in-spiralling object can penetrate deeper for a given core mass coordinate, thereby releasing more orbital energy and thus slightly facilitating
envelope ejection. See also Sect.~\ref{subsec:He-env} on extended envelopes of helium stars.}. 

\begin{figure}[t]
\begin{center}
\begin{minipage}{\columnwidth}
\begingroup
  \fontfamily{Hevetica}%
  \selectfont
  \makeatletter
  \providecommand\color[2][]{%
    \GenericError{(gnuplot) \space\space\space\@spaces}{%
      Package color not loaded in conjunction with
      terminal option `colourtext'%
    }{See the gnuplot documentation for explanation.%
    }{Either use 'blacktext' in gnuplot or load the package
      color.sty in LaTeX.}%
    \renewcommand\color[2][]{}%
  }%
  \providecommand\includegraphics[2][]{%
    \GenericError{(gnuplot) \space\space\space\@spaces}{%
      Package graphicx or graphics not loaded%
    }{See the gnuplot documentation for explanation.%
    }{The gnuplot epslatex terminal needs graphicx.sty or graphics.sty.}%
    \renewcommand\includegraphics[2][]{}%
  }%
  \providecommand\rotatebox[2]{#2}%
  \@ifundefined{ifGPcolor}{%
    \newif\ifGPcolor
    \GPcolortrue
  }{}%
  \@ifundefined{ifGPblacktext}{%
    \newif\ifGPblacktext
    \GPblacktexttrue
  }{}%
  \let\gplgaddtomacro\g@addto@macro
  \gdef\gplbacktext{}%
  \gdef\gplfronttext{}%
  \makeatother
  \ifGPblacktext
    \def\colorrgb#1{}%
    \def\colorgray#1{}%
  \else
    \ifGPcolor
      \def\colorrgb#1{\color[rgb]{#1}}%
      \def\colorgray#1{\color[gray]{#1}}%
      \expandafter\def\csname LTw\endcsname{\color{white}}%
      \expandafter\def\csname LTb\endcsname{\color{black}}%
      \expandafter\def\csname LTa\endcsname{\color{black}}%
      \expandafter\def\csname LT0\endcsname{\color[rgb]{1,0,0}}%
      \expandafter\def\csname LT1\endcsname{\color[rgb]{0,1,0}}%
      \expandafter\def\csname LT2\endcsname{\color[rgb]{0,0,1}}%
      \expandafter\def\csname LT3\endcsname{\color[rgb]{1,0,1}}%
      \expandafter\def\csname LT4\endcsname{\color[rgb]{0,1,1}}%
      \expandafter\def\csname LT5\endcsname{\color[rgb]{1,1,0}}%
      \expandafter\def\csname LT6\endcsname{\color[rgb]{0,0,0}}%
      \expandafter\def\csname LT7\endcsname{\color[rgb]{1,0.3,0}}%
      \expandafter\def\csname LT8\endcsname{\color[rgb]{0.5,0.5,0.5}}%
    \else
      \def\colorrgb#1{\color{black}}%
      \def\colorgray#1{\color[gray]{#1}}%
      \expandafter\def\csname LTw\endcsname{\color{white}}%
      \expandafter\def\csname LTb\endcsname{\color{black}}%
      \expandafter\def\csname LTa\endcsname{\color{black}}%
      \expandafter\def\csname LT0\endcsname{\color{black}}%
      \expandafter\def\csname LT1\endcsname{\color{black}}%
      \expandafter\def\csname LT2\endcsname{\color{black}}%
      \expandafter\def\csname LT3\endcsname{\color{black}}%
      \expandafter\def\csname LT4\endcsname{\color{black}}%
      \expandafter\def\csname LT5\endcsname{\color{black}}%
      \expandafter\def\csname LT6\endcsname{\color{black}}%
      \expandafter\def\csname LT7\endcsname{\color{black}}%
      \expandafter\def\csname LT8\endcsname{\color{black}}%
    \fi
  \fi
  \setlength{\unitlength}{0.0500bp}%
  \begin{picture}(5096.00,3822.00)%
    \gplgaddtomacro\gplbacktext{%
      \csname LTb\endcsname%
      \put(487,750){\makebox(0,0)[r]{\strut{}$10^{45}$}}%
      \put(487,1178){\makebox(0,0)[r]{\strut{}$10^{46}$}}%
      \put(487,1607){\makebox(0,0)[r]{\strut{}$10^{47}$}}%
      \put(487,2036){\makebox(0,0)[r]{\strut{}$10^{48}$}}%
      \put(487,2464){\makebox(0,0)[r]{\strut{}$10^{49}$}}%
      \put(487,2893){\makebox(0,0)[r]{\strut{}$10^{50}$}}%
      \put(487,3322){\makebox(0,0)[r]{\strut{}$10^{51}$}}%
      \put(1071,288){\makebox(0,0){\strut{}$1$}}%
      \put(2145,288){\makebox(0,0){\strut{}$10$}}%
      \put(3219,288){\makebox(0,0){\strut{}$100$}}%
      \put(4293,288){\makebox(0,0){\strut{}$1000$}}%
      \put(84,2090){\rotatebox{-270}{\makebox(0,0){\strut{}$-E~(\rm erg)$}}}%
      \put(2776,108){\makebox(0,0){\strut{}$\text{Radius coordinate,}~r~(\Rsun)$}}%
    }%
    \gplgaddtomacro\gplfronttext{%
      \csname LTb\endcsname%
      \put(4549,3432){\makebox(0,0)[r]{\strut{}convection}}%
      \csname LTb\endcsname%
      \put(4549,3162){\makebox(0,0)[r]{\strut{}$E_{\rm grav}$}}%
      \csname LTb\endcsname%
      \put(4549,2892){\makebox(0,0)[r]{\strut{}$\alpha_{\rm TH}=0.5$}}%
      \csname LTb\endcsname%
      \put(4549,2622){\makebox(0,0)[r]{\strut{}$E_{\rm bind}$}}%
    }%
    \gplgaddtomacro\gplbacktext{%
      \csname LTb\endcsname%
      \put(487,750){\makebox(0,0)[r]{\strut{}\phantom{$10^{45}$}}}%
      \put(487,1178){\makebox(0,0)[r]{\strut{}\phantom{$10^{46}$}}}%
      \put(487,1607){\makebox(0,0)[r]{\strut{}\phantom{$10^{47}$}}}%
      \put(487,2036){\makebox(0,0)[r]{\strut{}\phantom{$10^{48}$}}}%
      \put(487,2464){\makebox(0,0)[r]{\strut{}\phantom{$10^{49}$}}}%
      \put(487,2893){\makebox(0,0)[r]{\strut{}\phantom{$10^{50}$}}}%
      \put(487,3322){\makebox(0,0)[r]{\strut{}\phantom{$10^{51}$}}}%
      \put(1071,288){\makebox(0,0){\strut{}\phantom{$1$}}}%
      \put(2145,288){\makebox(0,0){\strut{}\phantom{$10$}}}%
      \put(3219,288){\makebox(0,0){\strut{}\phantom{$100$}}}%
      \put(4293,288){\makebox(0,0){\strut{}\phantom{$1000$}}}%
      \put(-24,2090){\rotatebox{-270}{\makebox(0,0){\strut{}\phantom{$-E~(\rm erg)$}}}}%
      \put(2776,108){\makebox(0,0){\strut{}\phantom{$\text{Radius coordinate,}~r~(\Rsun)$}}}%
    }%
    \gplgaddtomacro\gplfronttext{%
      \csname LTb\endcsname%
      \put(1208,2218){\makebox(0,0)[l]{\strut{}$\Delta E_{{\rm orb,}1.3}$}}%
      \csname LTb\endcsname%
      \put(1208,1948){\makebox(0,0)[l]{\strut{}$\Delta E_{{\rm orb,}5}$}}%
      \csname LTb\endcsname%
      \put(1208,1678){\makebox(0,0)[l]{\strut{}$\Delta E_{{\rm orb,}10}$}}%
      \csname LTb\endcsname%
      \put(1208,1408){\makebox(0,0)[l]{\strut{}$\Delta E_{{\rm orb,}20}$}}%
      \csname LTb\endcsname%
      \put(1208,1138){\makebox(0,0)[l]{\strut{}$\Delta E_{{\rm orb,}40}$}}%
      \csname LTb\endcsname%
      \put(1208,868){\makebox(0,0)[l]{\strut{}$\Delta E_{{\rm orb,}80}$}}%
    }%
    \gplgaddtomacro\gplbacktext{%
      \csname LTb\endcsname%
      \put(487,750){\makebox(0,0)[r]{\strut{}\phantom{$10^{45}$}}}%
      \put(487,1178){\makebox(0,0)[r]{\strut{}\phantom{$10^{46}$}}}%
      \put(487,1607){\makebox(0,0)[r]{\strut{}\phantom{$10^{47}$}}}%
      \put(487,2036){\makebox(0,0)[r]{\strut{}\phantom{$10^{48}$}}}%
      \put(487,2464){\makebox(0,0)[r]{\strut{}\phantom{$10^{49}$}}}%
      \put(487,2893){\makebox(0,0)[r]{\strut{}\phantom{$10^{50}$}}}%
      \put(487,3322){\makebox(0,0)[r]{\strut{}\phantom{$10^{51}$}}}%
      \put(1071,288){\makebox(0,0){\strut{}\phantom{$1$}}}%
      \put(2145,288){\makebox(0,0){\strut{}\phantom{$10$}}}%
      \put(3219,288){\makebox(0,0){\strut{}\phantom{$100$}}}%
      \put(4293,288){\makebox(0,0){\strut{}\phantom{$1000$}}}%
      \put(-24,2090){\rotatebox{-270}{\makebox(0,0){\strut{}\phantom{$-E~(\rm erg)$}}}}%
      \put(2776,108){\makebox(0,0){\strut{}\phantom{$\text{Radius coordinate,}~r~(\Rsun)$}}}%
    }%
    \gplgaddtomacro\gplfronttext{%
    }%
    \gplgaddtomacro\gplbacktext{%
      \csname LTb\endcsname%
      \put(487,750){\makebox(0,0)[r]{\strut{}\phantom{$10^{45}$}}}%
      \put(487,1178){\makebox(0,0)[r]{\strut{}\phantom{$10^{46}$}}}%
      \put(487,1607){\makebox(0,0)[r]{\strut{}\phantom{$10^{47}$}}}%
      \put(487,2036){\makebox(0,0)[r]{\strut{}\phantom{$10^{48}$}}}%
      \put(487,2464){\makebox(0,0)[r]{\strut{}\phantom{$10^{49}$}}}%
      \put(487,2893){\makebox(0,0)[r]{\strut{}\phantom{$10^{50}$}}}%
      \put(487,3322){\makebox(0,0)[r]{\strut{}\phantom{$10^{51}$}}}%
      \put(1071,288){\makebox(0,0){\strut{}\phantom{$1$}}}%
      \put(2145,288){\makebox(0,0){\strut{}\phantom{$10$}}}%
      \put(3219,288){\makebox(0,0){\strut{}\phantom{$100$}}}%
      \put(4293,288){\makebox(0,0){\strut{}\phantom{$1000$}}}%
      \put(-24,2090){\rotatebox{-270}{\makebox(0,0){\strut{}\phantom{$-E~(\rm erg)$}}}}%
      \put(2776,108){\makebox(0,0){\strut{}\phantom{$\text{Radius coordinate,}~r~(\Rsun)$}}}%
    }%
    \gplgaddtomacro\gplfronttext{%
    }%
    \gplbacktext
    \put(0,0){\includegraphics{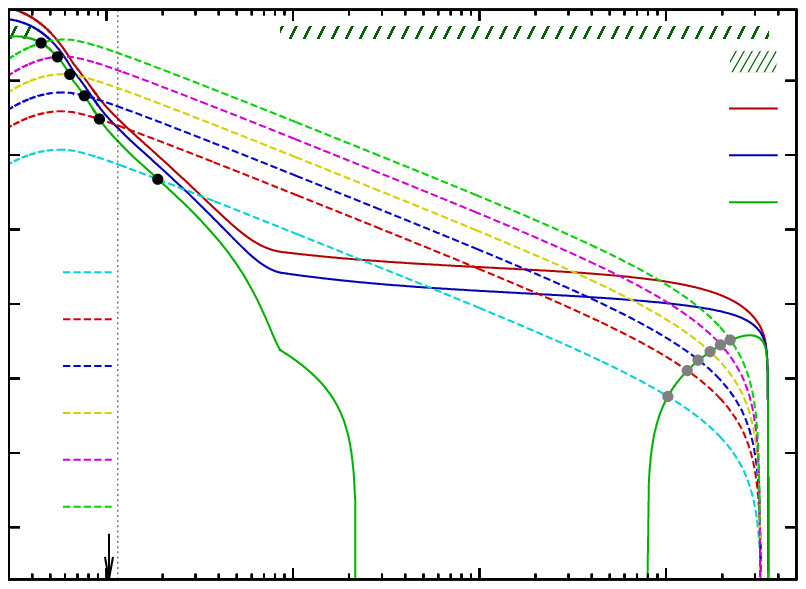}}%
    \gplfronttext
  \end{picture}%
\endgroup
\end{minipage}
  \caption[]{
Lower panel of Fig.~\ref{fig:structure} in radius coordinates. The additional blue solid line indicates where 50\% of the internal energy
is included in the calculated binding energy of the envelope. 
    }
\label{fig:structure_R}
\end{center}
\end{figure}
In the lower panel of Fig.~\ref{fig:structure}, the additional grey points mark the crossing where the released orbital energy exceeds the binding energy for the first time. 
Hence, an in-spiralling object has to spiral in at least to this depth to eject the material farther out. As long as there are no other energy sources, the in-spiral leading to
successful envelope ejection is expected to stop somewhere between the grey and the black points. 
Depending on the amount of the internal energy that can actually be used to eject the envelope \citep[see e.g.][for discussions]{hpe94,ijp15}, the crossing points should be be located 
somewhere between the solid green and red lines. 
The less the available internal energy, the deeper the in-spiral, and the more difficult it is for the binary system to eject the envelope (and survive instead of merging). 

It should be noted that the region between the two crossings of $\Delta E_{\rm orb}$ and $E_{\rm bind}$ shown in Fig.~\ref{fig:structure} has previously been discussed in terms of 'the energy expense' \citep{iva11}, that is,
the normalised excess energy available to the envelope after removal of all matter above a given mass coordinate.

\subsection{Response of donor star to mass loss}\label{subsec:response}
The immediate adiabatic response of the remaining envelope layers depends on whether they are convective or radiative \citep{hw87}.
On a longer timescale, the reaction of stripped cores to loss of their envelope depends on the amount of residual material remaining in the envelope.
The expansion or contraction of the remaining shell occurs on the thermal timescale of the remaining layer.
The residual hydrogen content following envelope ejection has been studied in the formation of WDs \citep{dv70}, and 
in particular in LMXB systems \citep{ts99,prp02,imt+16}. However, for more massive stars the situation is less well explored.

In Fig.~\ref{fig:structure_R} we plot the lower panel of Fig.~\ref{fig:structure} in radius coordinates. The convective envelope of this massive giant star spans a wide 
range in radius, although its mass is lower than the core mass. The core boundary is located near a radius coordinate of $r\simeq 1\;R_{\odot}$.
At first sight, the in-spiral might be expected to stop at the grey point, considering that at this location enough energy is released to unbind to material of the envelope farther out. 
However, when this point is well within hydrogen-rich material (which is clearly the case here), the system will not undergo final detachment at this location. 
As discussed above, the hydrogen-rich layers of the star will re-expand \citep[and during this thermal adjustment the star may develop strong thermal pulses, cf.][]{ijc+13}.
Hence, the mass transfer will continue and rebuild the CE. At this point, however, the drag force might be too weak to cause significant further in-spiral
before the core of the donor star collapses \citep[see e.g. Appendix~A in][for an estimate of the timescale of the in-spiral]{tlp15}. 
It is also possible that a self-regulated in-spiral is followed by additional RLO from either the core of the donor star (depending on the amount of residual hydrogen) or the in-spiralling companion star,
and this leads to further plunge-in. This plunge could in turn be followed by an additional self-regulated phase, and so on, such that a repetitive pattern may occur \citep{ijc+13}.
This pattern may repeat until the in-spiralling object reaches non-convective layers, at which point re-expansion of radiative material will process on a longer timescale.
In the case of our giant star model in Fig.~\ref{fig:structure_R}, the pre-CE convective boundary is located at a radius coordinate of about $8\;R_{\odot}$. 

The removal of the innermost hydrogen-rich layers may possibly proceed through dynamical stable mass transfer (this still has to be confirmed by numerical calculations),
until the mass of the diluted giant envelope reaches below a critical threshold value and the remaining envelope collapses and the binary finally becomes detached.
If mass is removed to below the bifurcation point, then the remaining core contracts on its thermal timescale. 

To summarise the above, we conclude that the termination point of the in-spiral is difficult to estimate accurately. We expect that at first, 
the plunge-in of the in-spiral will stop somewhere in the interval where the green curves bent vertically downward (i.e. where $E_{\rm bind}\ga 0$) 
in Fig.~\ref{fig:structure} (lower panel) and Fig.~\ref{fig:structure_R}.
The further evolution and the final post-CE orbital separation is not trivial to calculate and depends on the details of the physics of the
CE ejection process, the response of the remaining core to mass loss, and the amount of liberated accretion energy. 

We also conclude that the in-spiral will only come to an end and lead to successful CE ejection when both of the following conditions are fulfilled: 
\begin{itemize}
  \item The remaining amount of hydrogen is below the threshold for re-expansion of an envelope (i.e. the bifurcation point is located in a radiative layer with $X_H>0$).
  \vspace{0.3cm}
  \item The released orbital energy is sufficient to remove the envelope (i.e. the final location where in-spiral ends is between the black and grey points in 
         Figs.~\ref{fig:structure} and \ref{fig:structure_R}).
\end{itemize}
The last point illustrates once again the difficulty in population synthesis modelling of final post-CE orbital separations and thus explains 
the huge uncertainty in the LIGO merger rates determined from this method \citep{aaa+10}. Whereas the separation between the black and grey points for the massive giant star plotted
in Figs.~\ref{fig:structure} (lower panel) and~\ref{fig:structure_R} may cover a spread in mass coordinates of about $4\;M_{\odot}$ (less than 8\% of the remaining core mass), 
the corresponding spread in radius coordinates (and thus the spread in final post-CE orbital separation) is an astonishing factor 500!

It is therefore evident from our analysis that LIGO merger rates estimated from population synthesis of the CE formation channel \citep[e.g.][]{bkb02,vt03,bkr+08,dbf+12,mv14,bhbo16,es16}
must be highly uncertain and all quoted rates should be taken with a huge grain of salt (let alone other uncertain effects in addition to CE evolution).

\section{Discussions}\label{sec:discussions}
\subsection{Bifurcation point revisited}\label{subsec:bif_revisited}
A method suggested by \citet{iva11} is to locate the core boundary, after thermal readjustment, at the 
(pre-CE) mass coordinate in the hydrogen shell corresponding to the local maximum of the sonic velocity (i.e. at the maximum-compression point, $M_{\rm cp}$
where $P/\rho$ has a local maximum). 
It was argued that if a post-CE star has a final mass smaller than $M_{\rm cp}$ , then the star will shrink.
However, if it has any mass beyond this location, the star will continue to expand on the local thermal timescale. 
This may give rise to a new episode of mass transfer, or possibly a pulse. 

For our calculations of the core boundary in this study, we chose to use the $X_H=0.10$ criterion \citep{dt00}, which is often used in the literature.
Interestingly enough, we find that this point coincides closely to the maximum-compression point in the hydrogen shell burning layer in most of our models.
For all stellar models we investigated that are evolved beyond core hydrogen burning (independent of mass and metallicity), the locations of the core boundary
using the $X_H=0.10$ criterion and the maximum-compression point $M_{\rm cp}$ are often consistent to within 1\%, and always within 4\% 
(except for our few high-metallicity models with masses $<10\;M_{\odot}$, where the discrepancy can be up to 8\% in mass coordinate).

This general agreement is evident from comparing our $\lambda$-values, calculated with the $X_H=0.10$ criterion, with those recently calculated 
by \citet{wjl16b} for population~I stars of up to $60\;M_{\odot}$, using the $M_{\rm cp}$ criterion.

\subsection{Other energy sources}\label{subsec:other_sources}
According to \citet{ijc+13}, the question of additional energy sources, other than the release of orbital energy, depends partly on the extent to which the envelope is ejected directly by spiral shocks, 
developing from the orbital motion and tidal arms trailing the two stars \citep{rt12}, or indirectly by heating and a pressure gradient.
If the donor star core expands as a consequence of mass loss, it could do mechanical work on the envelope and change the boundary conditions
for the integral in Eq.~(\ref{eq:bindingenergy}).
Enforcing corotation of the envelope through tidal heating may produce an energy sink.

\citet{ijp15} demonstrated that heat input leading to kinetic energy deposition within the envelope is not just a simple function of
radius and mass. It depends on the structure of the pre-CE donor star (e.g. mass density profile and the degree of corotation), the initial mass ratio
between the two stars, and on how angular momentum is transported through the CE. In other words, the authors concluded that the envelope ejection process depends 
on i) the amount, ii) the location,  and iii) how rapidly the released energy is transferred to the envelope, and they predict two types of outcomes: 
'runaway' and 'self-regulated' envelope ejection.

\subsubsection{Recombination energy}
The inclusion of recombination energy of hydrogen and helium \citep[e.g.][]{hpe94} 
has been argued to be a promising candidate for producing successful envelope ejection \citep[e.g.][and references therein]{ijp15}.
Recent 3D hydrodynamical modelling \citep{nil15} taking the released recombination energy reservoir into account, let to the first successful CE ejection and production
of a post-CE double WD system. 

\begin{figure}[t]
\begin{center}
\begin{minipage}{\columnwidth}
\begingroup
  \fontfamily{Hevetica}%
  \selectfont
  \makeatletter
  \providecommand\color[2][]{%
    \GenericError{(gnuplot) \space\space\space\@spaces}{%
      Package color not loaded in conjunction with
      terminal option `colourtext'%
    }{See the gnuplot documentation for explanation.%
    }{Either use 'blacktext' in gnuplot or load the package
      color.sty in LaTeX.}%
    \renewcommand\color[2][]{}%
  }%
  \providecommand\includegraphics[2][]{%
    \GenericError{(gnuplot) \space\space\space\@spaces}{%
      Package graphicx or graphics not loaded%
    }{See the gnuplot documentation for explanation.%
    }{The gnuplot epslatex terminal needs graphicx.sty or graphics.sty.}%
    \renewcommand\includegraphics[2][]{}%
  }%
  \providecommand\rotatebox[2]{#2}%
  \@ifundefined{ifGPcolor}{%
    \newif\ifGPcolor
    \GPcolortrue
  }{}%
  \@ifundefined{ifGPblacktext}{%
    \newif\ifGPblacktext
    \GPblacktexttrue
  }{}%
  \let\gplgaddtomacro\g@addto@macro
  \gdef\gplbacktext{}%
  \gdef\gplfronttext{}%
  \makeatother
  \ifGPblacktext
    \def\colorrgb#1{}%
    \def\colorgray#1{}%
  \else
    \ifGPcolor
      \def\colorrgb#1{\color[rgb]{#1}}%
      \def\colorgray#1{\color[gray]{#1}}%
      \expandafter\def\csname LTw\endcsname{\color{white}}%
      \expandafter\def\csname LTb\endcsname{\color{black}}%
      \expandafter\def\csname LTa\endcsname{\color{black}}%
      \expandafter\def\csname LT0\endcsname{\color[rgb]{1,0,0}}%
      \expandafter\def\csname LT1\endcsname{\color[rgb]{0,1,0}}%
      \expandafter\def\csname LT2\endcsname{\color[rgb]{0,0,1}}%
      \expandafter\def\csname LT3\endcsname{\color[rgb]{1,0,1}}%
      \expandafter\def\csname LT4\endcsname{\color[rgb]{0,1,1}}%
      \expandafter\def\csname LT5\endcsname{\color[rgb]{1,1,0}}%
      \expandafter\def\csname LT6\endcsname{\color[rgb]{0,0,0}}%
      \expandafter\def\csname LT7\endcsname{\color[rgb]{1,0.3,0}}%
      \expandafter\def\csname LT8\endcsname{\color[rgb]{0.5,0.5,0.5}}%
    \else
      \def\colorrgb#1{\color{black}}%
      \def\colorgray#1{\color[gray]{#1}}%
      \expandafter\def\csname LTw\endcsname{\color{white}}%
      \expandafter\def\csname LTb\endcsname{\color{black}}%
      \expandafter\def\csname LTa\endcsname{\color{black}}%
      \expandafter\def\csname LT0\endcsname{\color{black}}%
      \expandafter\def\csname LT1\endcsname{\color{black}}%
      \expandafter\def\csname LT2\endcsname{\color{black}}%
      \expandafter\def\csname LT3\endcsname{\color{black}}%
      \expandafter\def\csname LT4\endcsname{\color{black}}%
      \expandafter\def\csname LT5\endcsname{\color{black}}%
      \expandafter\def\csname LT6\endcsname{\color{black}}%
      \expandafter\def\csname LT7\endcsname{\color{black}}%
      \expandafter\def\csname LT8\endcsname{\color{black}}%
    \fi
  \fi
  \setlength{\unitlength}{0.0500bp}%
  \begin{picture}(5096.00,3822.00)%
    \gplgaddtomacro\gplbacktext{%
      \csname LTb\endcsname%
      \put(487,1351){\makebox(0,0)[r]{\strut{}$0.01$}}%
      \put(487,2641){\makebox(0,0)[r]{\strut{}$0.1$}}%
      \put(1809,288){\makebox(0,0){\strut{}$10$}}%
      \put(4296,288){\makebox(0,0){\strut{}$100$}}%
      \put(84,2090){\rotatebox{-270}{\makebox(0,0){\strut{}$E_{\rm recombination}/U$}}}%
      \put(2776,108){\makebox(0,0){\strut{}$M_{\rm ZAMS}$ in $\Msun$}}%
    }%
    \gplgaddtomacro\gplfronttext{%
      \csname LTb\endcsname%
      \put(758,614){\makebox(0,0)[l]{\strut{}MW}}%
      \put(758,877){\makebox(0,0)[l]{\strut{}IZw18}}%
      \put(758,1139){\makebox(0,0)[l]{\strut{}IZw18 $M_{\rm core}^{\rm H}=1\,\Msun$}}%
    }%
    \gplbacktext
    \put(0,0){\includegraphics{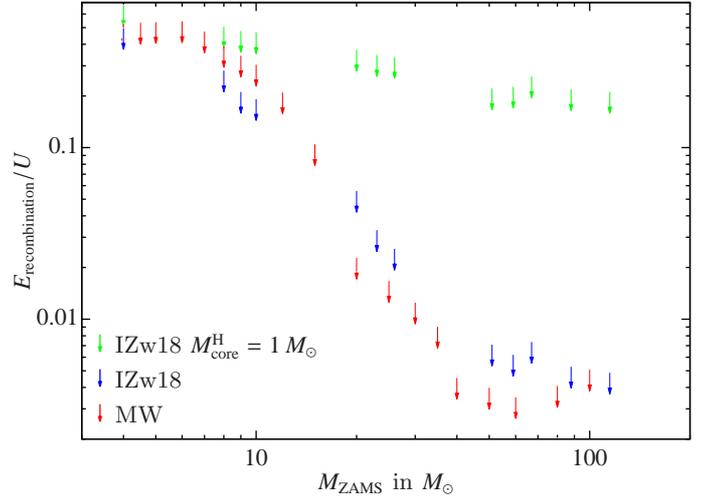}}%
    \gplfronttext
  \end{picture}%
\endgroup
\end{minipage}
  \caption[]{
     Recombination energy in the envelope as a function of ZAMS mass. The plot shows the maximum contribution of recombination energy (from H, He and H$_2$) to the total internal energy ($U$)
     obtained during the evolution of the stars. These maximum values are always reached when the stars are near their largest radial expansion as giants. 
     The red arrows are for $Z=Z_{\rm MW}$ and the blue ones for $Z=Z_{\odot}/50$. The green arrows are also calculated for our $Z=Z_{\odot}/50$ models, but assume a larger remaining core
     that includes $1\;M_{\odot}$ of hydrogen. 
    }
\label{fig:recomb}
\end{center}
\end{figure}

Figure~\ref{fig:recomb} shows the significance of recombination energy in units of the total internal energy $U$ in the envelopes of our stellar models. 
Within the core of the star, the internal energy is fully dominated by radiation and gas pressure. The recombination energy contributes strongest to the total 
internal energy in the outer regions of the envelope with a mass density inversion.

Whereas the recombination energy can be an important contribution (up to $\sim\!55$\% of $U$) for low- and intermediate-mass stars, it does not play a role for 
massive stars when we apply the $X_{\rm H}=0.10$ criterion for the core boundary (red and blue arrows). 
For BH progenitors, the contribution is typically lower than 1\%, which may potentially lead to problems using current hydrodynamical simulation codes  
because they apparently only succeed to eject the envelope of low-mass stars when taking the released recombination energy into account \citep{nil15}. 
As Fig.~\ref{fig:recomb} points out, the metallicity content has no significant effect on this general behaviour.

Changing the bifurcation point criterion, however, such that the remaining core is assumed to include $1\;M_{\odot}$ of hydrogen, causes the relative contribution of
recombination energy to be more important (green arrows) and thus play a role in facilitating CE ejection.
An additional effect that favours successful ejection of CEs in wide systems is that the released recombination energy is highest for the most extended (coldest) stars.

\subsubsection{Enthalpy}\label{subsubsec:enthalpy}
\citet{ic11} argued that including the enthalpy in the energy budget typically results in $\lambda$-values that are higher by a factor of 2 to 3 \citep[see also][]{wjl16b}.
Whether enthalpy should be included in the CE energy budget at all is controversial and may depend on the timescale of the CE ejection. 
Rather than being a new energy source, the main contribution of the $P/\rho$ term is that it redistributes energy:  
it adds more kinetic energy to the gas ejected from the outer envelope regions at the expense of the energy
of the inner regions of the envelope \citep{ijc+13}. This may cause the formation of a circumbinary disk if the
inner envelope material is barely ejected at the escape velocity. Such a circumbinary disk can act as an additional
sink of orbital angular momentum losses \citep[e.g.][]{spv97,st01}.
As a result of the dispute and uncertainty of potentially including the $P/\rho$ term in Eq.~\eqref{eq:bindingenergy}, we disregard this term in our modelling. 

\subsubsection{Liberated accretion energy}\label{subsubsec:acc_energy}
Release of accretion energy is an additional energy source. This contribution may even dominate that of orbital energy release
in the beginning of the in-spiral, for the reason that in the outer envelope layers the in-spiral timescale is relatively long and the binding energy
per unit mass is low. 
In recent studies of hydrodynamical simulations, \citet{mr15a,mr15b} found that a compact object such as a NS
embedded in a CE only accretes a very modest amount of material during its in-spiral as a result of a density gradient across its
accretion radius, which strongly limits accretion by imposing a net angular momentum to the flow around the NS.
This conclusion supports earlier work by \citet{rt12}, who also found that the true accretion rate of the accreting star
is much lower than predicted by the Bondi-Hoyle prescription.
Nevertheless, even modest accretion rates constrained by the Eddington limit can contribute significant heat to the CE energy budget.

\citet{vt03} introduced in their simulations the inclusion of released accretion energy from the in-spiralling
compact object (thus facilitating envelope ejection), and hence demonstrated that the expected aLIGO detection rate of BH-BH mergers should strongly dominate that of NS-NS systems. The energy input from accretion onto a BH during a CE phase is given by
$\Delta E_{\rm acc}=\eta\,\dot{M}_{\rm Edd}c^2\,\tau _{\rm CE}$, where the Eddington accretion limit \citep{vdh94a} can be estimated as
\begin{equation}
  \dot{M}_{\rm Edd} = 4.4\times 10^{-9}\;M_{\odot}\,{\rm yr}^{-1}\;\left(\frac{M_{\rm BH}}{M_{\odot}}\right)\,\frac{r_{\ast}}{(1+X_H)}\;,
\end{equation}
yielding
\begin{equation}
  \Delta E_{\rm acc} = 1.6\times 10^{48}\;{\rm erg} \;\left(\frac{M_{\rm BH}}{M_{\odot}}\right)\,\left(\frac{\tau_{\rm CE}}{{\rm 1000~yr}}\right)\,\left(\frac{\eta}{{0.20}}\right)\,\frac{r_{\ast}}{(1+X_H)}\;.
\end{equation}
Here, $\tau _{\rm CE}<10^3\;{\rm yr}$ is the duration of the CE phase \citep[dictated by the thermal timescale of the envelope,][]{ijc+13}, 
$X_H$ is the mass fraction of hydrogen in the donor-star envelope, $\eta$ is the accretion 
radiation efficiency, and $r_{\ast}=R_{\rm ISCO}/(GM_{\rm BH}/c^2)$ is the location of the innermost stable circular orbit (ISCO).
Both $\eta=0.06-0.42$ and $r_{\ast}=1-6$ depend on the (here assumed to be prograde) spin of the accreting BH. 

As an example, a $35\;M_{\odot}$~BH with an Eddington-limited accretion rate ($\sim\! 10^{-7}\;M_{\odot}\,{\rm yr}^{-1}$) 
would therefore be able to accrete about $10^{-4}\;M_{\odot}$ while embedded in a CE and release a total energy output of
$\sim\!5\times 10^{49}\;{\rm erg}$, which can potentially be used to eject the envelope. As can be seen from Fig.~\ref{fig:structure_R}, a heat input of this amount
could significantly facilitate envelope ejection even in massive stars. However, as we discuss
in Sect.~\ref{subsubsec:energy_transport} below, this possibility depends on the physics of energy transport in the envelope to be ejected.

In addition to heating, we note that accretion can also help in envelope ejection by the possible formation of a jet by the in-spiralling object \citep{sok04,sok16}. 
Especially BHs and NSs are expected to potentially launch very energetic jets.
In a scenario recently suggested by \citet{sok15}, so-called grazing envelope evolution might be made possible if a compact companion
star manages to accrete matter at a high rate and launch a jet that removes the outskirts of the giant envelope, 
hence preventing the formation of a CE. However, further investigation of this model is needed.

\subsubsection{Convective energy transport}\label{subsubsec:energy_transport}
In order to eject a CE, the liberated energy, either from the in-spiral of the compact companion, from accretion onto this compact companion, 
or from the recombination of ionised envelope material needs to be converted into mechanical energy. If all these processes take
place inside a fully convective envelope, the question arises whether a part of the liberated energy, which is at first present in the form
of heat, would be quickly transported to the top of the envelope, where it would be radiated away.

The efficiency of this energy loss will depend on the ratio of the timescale of energy liberation to the convective timescale.
If it is small, then convective energy loss will be negligible. If the ratio is near one or higher, convective energy loss may be important.
As the convective timescale is of the order of the dynamical timescale of the star, it appears possible that convective energy loss 
is relevant for all three forms of energy liberation mentioned above. It will require models of time-dependent convection to quantify this effect.

\subsection{Ejection efficiency parameter}\label{subsec:alpha}
So far, we have not addressed the value of the ejection efficiency parameter, which we have simply assumed to be $\alpha=1$.
There are several reasons, however, why a realistic value of the ejection efficiency parameter would be $\alpha < 1$. 
An example is radiative losses from the CE (e.g. as discussed above in Sect.~\ref{subsubsec:energy_transport}) or internal and kinetic energy of the ejecta material.

Energy loss from the envelope photosphere is relevant for relatively slow, thermal timescale CE events (in which case there
might also be significant energy input from the naked, hot stellar core). Moreover, recent work by \citet{nil15} demonstrated a
case where between 25\% to 50\% of the released orbital energy is taken away as kinetic energy of the ejected material, implying $\alpha < 0.75$ from this effect alone.

Assuming lower and more realistic values of $\alpha$ (e.g. $0.3-0.7$) would cause all the solid lines in Fig.~\ref{fig:Mxmin} to move up, 
and all dashed lines in Figs.~\ref{fig:structure} and \ref{fig:structure_R} to move down. 
For example, in Figs.~\ref{fig:structure} and \ref{fig:structure_R}, the line of the $5\;M_{\odot}$ in-spiralling object for $\alpha=1$ is 
comparable to that of a $10\;M_{\odot}$ in-spiralling object with $\alpha\approx 0.5$. 

\subsection{Post-CE orbital separations in population synthesis}\label{subsec:pop_syn}
\begin{figure}[t]
\begin{center}
\begin{minipage}{\columnwidth}
\begingroup
  \fontfamily{Hevetica}%
  \selectfont
  \makeatletter
  \providecommand\color[2][]{%
    \GenericError{(gnuplot) \space\space\space\@spaces}{%
      Package color not loaded in conjunction with
      terminal option `colourtext'%
    }{See the gnuplot documentation for explanation.%
    }{Either use 'blacktext' in gnuplot or load the package
      color.sty in LaTeX.}%
    \renewcommand\color[2][]{}%
  }%
  \providecommand\includegraphics[2][]{%
    \GenericError{(gnuplot) \space\space\space\@spaces}{%
      Package graphicx or graphics not loaded%
    }{See the gnuplot documentation for explanation.%
    }{The gnuplot epslatex terminal needs graphicx.sty or graphics.sty.}%
    \renewcommand\includegraphics[2][]{}%
  }%
  \providecommand\rotatebox[2]{#2}%
  \@ifundefined{ifGPcolor}{%
    \newif\ifGPcolor
    \GPcolortrue
  }{}%
  \@ifundefined{ifGPblacktext}{%
    \newif\ifGPblacktext
    \GPblacktexttrue
  }{}%
  \let\gplgaddtomacro\g@addto@macro
  \gdef\gplbacktext{}%
  \gdef\gplfronttext{}%
  \makeatother
  \ifGPblacktext
    \def\colorrgb#1{}%
    \def\colorgray#1{}%
  \else
    \ifGPcolor
      \def\colorrgb#1{\color[rgb]{#1}}%
      \def\colorgray#1{\color[gray]{#1}}%
      \expandafter\def\csname LTw\endcsname{\color{white}}%
      \expandafter\def\csname LTb\endcsname{\color{black}}%
      \expandafter\def\csname LTa\endcsname{\color{black}}%
      \expandafter\def\csname LT0\endcsname{\color[rgb]{1,0,0}}%
      \expandafter\def\csname LT1\endcsname{\color[rgb]{0,1,0}}%
      \expandafter\def\csname LT2\endcsname{\color[rgb]{0,0,1}}%
      \expandafter\def\csname LT3\endcsname{\color[rgb]{1,0,1}}%
      \expandafter\def\csname LT4\endcsname{\color[rgb]{0,1,1}}%
      \expandafter\def\csname LT5\endcsname{\color[rgb]{1,1,0}}%
      \expandafter\def\csname LT6\endcsname{\color[rgb]{0,0,0}}%
      \expandafter\def\csname LT7\endcsname{\color[rgb]{1,0.3,0}}%
      \expandafter\def\csname LT8\endcsname{\color[rgb]{0.5,0.5,0.5}}%
    \else
      \def\colorrgb#1{\color{black}}%
      \def\colorgray#1{\color[gray]{#1}}%
      \expandafter\def\csname LTw\endcsname{\color{white}}%
      \expandafter\def\csname LTb\endcsname{\color{black}}%
      \expandafter\def\csname LTa\endcsname{\color{black}}%
      \expandafter\def\csname LT0\endcsname{\color{black}}%
      \expandafter\def\csname LT1\endcsname{\color{black}}%
      \expandafter\def\csname LT2\endcsname{\color{black}}%
      \expandafter\def\csname LT3\endcsname{\color{black}}%
      \expandafter\def\csname LT4\endcsname{\color{black}}%
      \expandafter\def\csname LT5\endcsname{\color{black}}%
      \expandafter\def\csname LT6\endcsname{\color{black}}%
      \expandafter\def\csname LT7\endcsname{\color{black}}%
      \expandafter\def\csname LT8\endcsname{\color{black}}%
    \fi
  \fi
  \setlength{\unitlength}{0.0500bp}%
  \begin{picture}(5096.00,3822.00)%
    \gplgaddtomacro\gplbacktext{%
    }%
    \gplgaddtomacro\gplfronttext{%
      \csname LTb\endcsname%
      \put(4549,905){\makebox(0,0)[r]{\strut{}$M_{\rm X}=80\,\Msun$}}%
      \csname LTb\endcsname%
      \put(4549,635){\makebox(0,0)[r]{\strut{}$M_{\rm X}=35\,\Msun$}}%
    }%
    \gplgaddtomacro\gplbacktext{%
    }%
    \gplgaddtomacro\gplfronttext{%
      \csname LTb\endcsname%
      \put(3143,905){\makebox(0,0)[r]{\strut{}$M_{\rm X}=10\,\Msun$}}%
      \csname LTb\endcsname%
      \put(3143,635){\makebox(0,0)[r]{\strut{}$M_{\rm X}=\enspace5\,\Msun$}}%
    }%
    \gplgaddtomacro\gplbacktext{%
      \csname LTb\endcsname%
      \put(334,1314){\makebox(0,0)[r]{\strut{}$1$}}%
      \put(334,2551){\makebox(0,0)[r]{\strut{}$10$}}%
      \put(356,288){\makebox(0,0){\strut{}$10$}}%
      \put(1919,288){\makebox(0,0){\strut{}$10^{2}$}}%
      \put(3481,288){\makebox(0,0){\strut{}$10^{3}$}}%
      \put(5044,288){\makebox(0,0){\strut{}$10^{4}$}}%
      \put(82,2090){\rotatebox{-270}{\makebox(0,0){\strut{}$\text{post-CE separation,}~a_{\rm f}~(\Rsun)$}}}%
      \put(2700,108){\makebox(0,0){\strut{}$\text{pre-CE separation,}~a_{\rm i}~(\Rsun)$}}%
      \put(631,3415){\makebox(0,0)[l]{\strut{}GW}}%
      \put(631,2179){\makebox(0,0)[l]{\strut{}CE}}%
    }%
    \gplgaddtomacro\gplfronttext{%
    }%
    \gplbacktext
    \put(0,0){\includegraphics{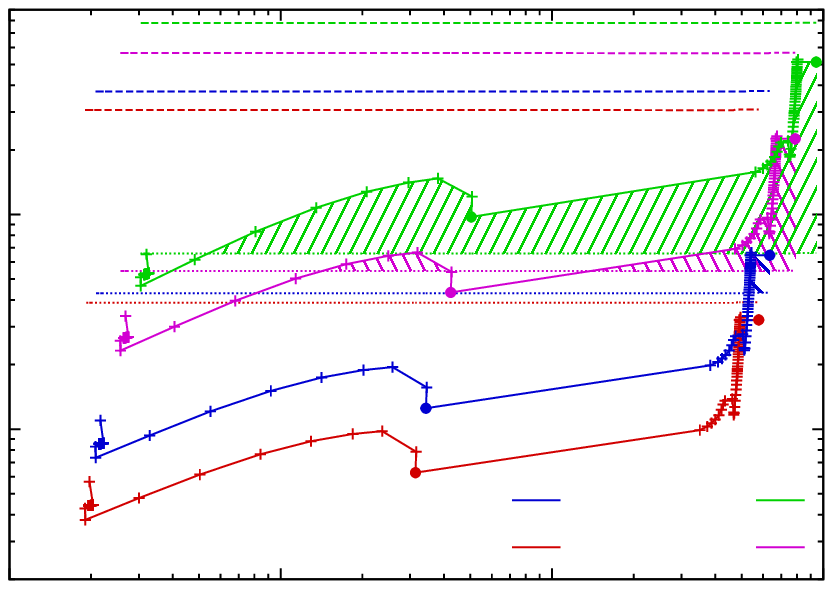}}%
    \gplfronttext
  \end{picture}%
\endgroup

\begingroup
  \fontfamily{Hevetica}%
  \selectfont
  \makeatletter
  \providecommand\color[2][]{%
    \GenericError{(gnuplot) \space\space\space\@spaces}{%
      Package color not loaded in conjunction with
      terminal option `colourtext'%
    }{See the gnuplot documentation for explanation.%
    }{Either use 'blacktext' in gnuplot or load the package
      color.sty in LaTeX.}%
    \renewcommand\color[2][]{}%
  }%
  \providecommand\includegraphics[2][]{%
    \GenericError{(gnuplot) \space\space\space\@spaces}{%
      Package graphicx or graphics not loaded%
    }{See the gnuplot documentation for explanation.%
    }{The gnuplot epslatex terminal needs graphicx.sty or graphics.sty.}%
    \renewcommand\includegraphics[2][]{}%
  }%
  \providecommand\rotatebox[2]{#2}%
  \@ifundefined{ifGPcolor}{%
    \newif\ifGPcolor
    \GPcolortrue
  }{}%
  \@ifundefined{ifGPblacktext}{%
    \newif\ifGPblacktext
    \GPblacktexttrue
  }{}%
  \let\gplgaddtomacro\g@addto@macro
  \gdef\gplbacktext{}%
  \gdef\gplfronttext{}%
  \makeatother
  \ifGPblacktext
    \def\colorrgb#1{}%
    \def\colorgray#1{}%
  \else
    \ifGPcolor
      \def\colorrgb#1{\color[rgb]{#1}}%
      \def\colorgray#1{\color[gray]{#1}}%
      \expandafter\def\csname LTw\endcsname{\color{white}}%
      \expandafter\def\csname LTb\endcsname{\color{black}}%
      \expandafter\def\csname LTa\endcsname{\color{black}}%
      \expandafter\def\csname LT0\endcsname{\color[rgb]{1,0,0}}%
      \expandafter\def\csname LT1\endcsname{\color[rgb]{0,1,0}}%
      \expandafter\def\csname LT2\endcsname{\color[rgb]{0,0,1}}%
      \expandafter\def\csname LT3\endcsname{\color[rgb]{1,0,1}}%
      \expandafter\def\csname LT4\endcsname{\color[rgb]{0,1,1}}%
      \expandafter\def\csname LT5\endcsname{\color[rgb]{1,1,0}}%
      \expandafter\def\csname LT6\endcsname{\color[rgb]{0,0,0}}%
      \expandafter\def\csname LT7\endcsname{\color[rgb]{1,0.3,0}}%
      \expandafter\def\csname LT8\endcsname{\color[rgb]{0.5,0.5,0.5}}%
    \else
      \def\colorrgb#1{\color{black}}%
      \def\colorgray#1{\color[gray]{#1}}%
      \expandafter\def\csname LTw\endcsname{\color{white}}%
      \expandafter\def\csname LTb\endcsname{\color{black}}%
      \expandafter\def\csname LTa\endcsname{\color{black}}%
      \expandafter\def\csname LT0\endcsname{\color{black}}%
      \expandafter\def\csname LT1\endcsname{\color{black}}%
      \expandafter\def\csname LT2\endcsname{\color{black}}%
      \expandafter\def\csname LT3\endcsname{\color{black}}%
      \expandafter\def\csname LT4\endcsname{\color{black}}%
      \expandafter\def\csname LT5\endcsname{\color{black}}%
      \expandafter\def\csname LT6\endcsname{\color{black}}%
      \expandafter\def\csname LT7\endcsname{\color{black}}%
      \expandafter\def\csname LT8\endcsname{\color{black}}%
    \fi
  \fi
  \setlength{\unitlength}{0.0500bp}%
  \begin{picture}(5096.00,3822.00)%
    \gplgaddtomacro\gplbacktext{%
    }%
    \gplgaddtomacro\gplfronttext{%
      \csname LTb\endcsname%
      \put(4549,1155){\makebox(0,0)[r]{\strut{}acc. energy}}%
    }%
    \gplgaddtomacro\gplbacktext{%
    }%
    \gplgaddtomacro\gplfronttext{%
      \csname LTb\endcsname%
      \put(4549,897){\makebox(0,0)[r]{\strut{}$\alpha=100\%$}}%
      \csname LTb\endcsname%
      \put(4549,627){\makebox(0,0)[r]{\strut{}$\alpha=\enspace75\%$}}%
    }%
    \gplgaddtomacro\gplbacktext{%
    }%
    \gplgaddtomacro\gplfronttext{%
      \csname LTb\endcsname%
      \put(3143,897){\makebox(0,0)[r]{\strut{}$\alpha=\enspace50\%$}}%
      \csname LTb\endcsname%
      \put(3143,627){\makebox(0,0)[r]{\strut{}$\alpha=\enspace25\%$}}%
    }%
    \gplgaddtomacro\gplbacktext{%
      \csname LTb\endcsname%
      \put(334,1032){\makebox(0,0)[r]{\strut{}$1$}}%
      \put(334,2495){\makebox(0,0)[r]{\strut{}$10$}}%
      \put(1591,288){\makebox(0,0){\strut{}$10^{2}$}}%
      \put(3358,288){\makebox(0,0){\strut{}$10^{3}$}}%
      \put(82,2090){\rotatebox{-270}{\makebox(0,0){\strut{}$\text{post-CE separation,}~a_{\rm f}~(\Rsun)$}}}%
      \put(2700,108){\makebox(0,0){\strut{}$\text{pre-CE separation,}~a_{\rm i}~(\Rsun)$}}%
      \put(527,3375){\makebox(0,0)[l]{\strut{}GW-merger limit}}%
      \put(527,2268){\makebox(0,0)[l]{\strut{}CE-merger limit}}%
    }%
    \gplgaddtomacro\gplfronttext{%
    }%
    \gplbacktext
    \put(0,0){\includegraphics{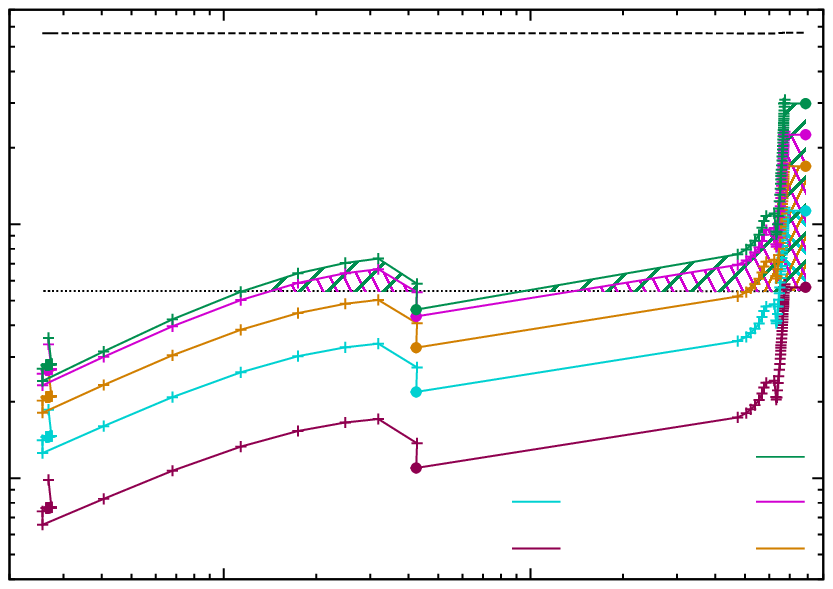}}%
    \gplfronttext
  \end{picture}%
\endgroup
\end{minipage}
  \caption[]{Mapping of pre-CE to post-CE orbital separations as done in population synthesis (see Sect.~\ref{subsec:pop_syn}) when applying the $(\alpha,\lambda)$-formalism.
             The donor is assumed to be our $88\;M_{\odot}$ ($Z=Z_{\odot}/50$) star. The in-spiralling objects (BHs) have masses of between 5 and $80\;M_{\odot}$.
             Below the dashed GW lines, the post-CE systems will merge within a Hubble time. Below the dotted CE lines, the in-spiral continues to the core of the donor star 
             and leads to coalescence (i.e. the relaxed He-core fills its Roche-lobe, cf. Sect.~\ref{subsec:He-env}, and the binary will not survive). 
             In the upper panel, the different colours mark the mass of the in-spiralling object (using $\alpha=1$); in the lower panel, they represent different values of $\alpha$
             (for a fixed value of $M_{\rm X}=35\;\Msun$). The dark green (upper) line in the lower panel was calculated for $\alpha=1$ and an additional accretion energy of $5\times 10^{49}\;{\rm erg}$, 
             see Sect.~\ref{subsubsec:acc_energy}. The hatched regions indicate systems that are expected to successfully produce LIGO sources.
             The solid dots mark the models shown in Figs.~\ref{fig:structure} and \ref{fig:structure_R}.
    }
\label{fig:pop_syn}
\end{center}
\end{figure}
For a discussion of predicted LIGO detection rates of merging BH-BH, NS-NS, or BH-NS binaries, it is of interest to evaluate the amount of fine-tuning needed for 
a given binary system to survive CE evolution, and to probe how the mapping of pre-CE orbital separations to post-CE orbital separations are performed in a typical population synthesis code. 
In such codes, it is usually assumed that all material is removed above a core boundary at $X_H=0.10$. In the discussion below we therefore
apply this bifurcation point criterion. 

In Fig.~\ref{fig:pop_syn} we plot post-CE orbital separations, $a_{\rm f}$ as a function of pre-CE orbital separations, $a_{\rm i}$ for the $88\;M_{\odot}$ ($Z=Z_{\odot}/50$) donor star
investigated in this paper. In the upper panel, we show the results for in-spiralling companions (BHs) of masses: 5, 10, 35 and $80\;M_{\odot}$, in all cases assuming 
an envelope ejection efficiency parameter of $\alpha=1$.
In the lower panel, we assume $M_{\rm X}=35\;M_{\odot}$ (i.e. resembling the progenitor system of GW150914) for different values of $\alpha$ (0.25, 0.50, 0.75 and 1),
and including one additional case (for $\alpha=1$) where we assumed injection of released accretion energy of $\Delta E_{\rm acc}=5\times10^{49}\;{\rm erg}$.

In each panel we show GW and CE lines, corresponding to post-CE orbital separations below which the system will merge within a Hubble time (13.8~Gyr) and thus become
detectable as a gravitational wave source, or coalesce during the CE in-spiral and thus not survive as a binary system, respectively.  
For each system, the intervals of $a_{\rm i}$, for which the binary successfully survives and eventually produces a LIGO merger event within a Hubble time,
are marked with a hatched pattern. For example, the upper panel shows that only the two in-spiralling objects with masses $M_{\rm X}=80\;M_{\odot}$ and 
$M_{\rm X}=35\;M_{\odot}$ can successfully eject the CE of our donor star (which has an initial mass of $88\;M_{\odot}$ and $Z=Z_{\odot}/50$) before it reaches its giant stage.
The in-spiralling object with $M_{\rm X}=10\;M_{\odot}$ is only able to eject the envelope of the donor star when the latter has evolved to its very last expansion phase as a giant. 
The in-spiralling $5\;M_{\odot}$ cannot eject the envelope at all (see also Fig.~\ref{fig:Mxmin}).
For $M_{\rm X}>27\;M_{\odot}$, there are two windows of $a_{\rm i}$ intervals that allow CE ejection.

Whereas for a massive in-spiralling object of $80\;M_{\odot}$ we find solutions for the entire interval $60\le a_{\rm i}<10\,000\;R_{\odot}$,
an in-spiralling object with $M_{\rm X}=35\;M_{\odot}$ only has solutions for $130< a_{\rm i}<400\;R_{\odot}$ and $1300< a_{\rm i}<8000\;R_{\odot}$.
Given their lower values of $a_{\rm f}$, the $M_{\rm X}=35\;M_{\odot}$ systems produce shorter delay-time binaries (i.e. they merge on a shorter timescale following the CE ejection
than the $M_{\rm X}=80\;M_{\odot}$ systems). 

In the lower panel of Fig.~\ref{fig:pop_syn}, an envelope ejection efficiency close to 100\% ($\alpha=1$) is needed for the
system with $M_{\rm X}=35\;M_{\odot}$ to survive. An injection of $\Delta E_{\rm acc}=5\times 10^{49}\;{\rm erg}$ will result in somewhat less in-spiral
and therefore wider post-CE binaries with longer delay-times. 

The points on each curve in Fig.~\ref{fig:pop_syn} were calculated using a specific subroutine of the binary population synthesis code of Kruckow~et~al. (in~prep.),
which estimates the post-CE orbital separations. This code makes use of interpolations of stellar tracks using a finite
grid resolution of stellar radii 
that is combined with the dimensionless Roche-lobe radius \citep{egg83} to determine $a_{\rm i}$ for each value of $M_{\rm X}$. The values of $a_{\rm f}$
are then determined from tabulated values of $\lambda$ associated with the stellar grids, following Eq.(\ref{eq:bindingenergy}) and combined with 
$|E_{\rm bind}|=\alpha \cdot |\Delta E_{\rm orb}| + \Delta E_{\rm acc}$. 

\subsection{Post-CE merger before core collapse?}\label{subsec:preSN_merger}
\begin{figure}[t]
\begin{center}
\vspace{0.2cm}
 \mbox{\includegraphics[width=0.35\textwidth, angle=-90]{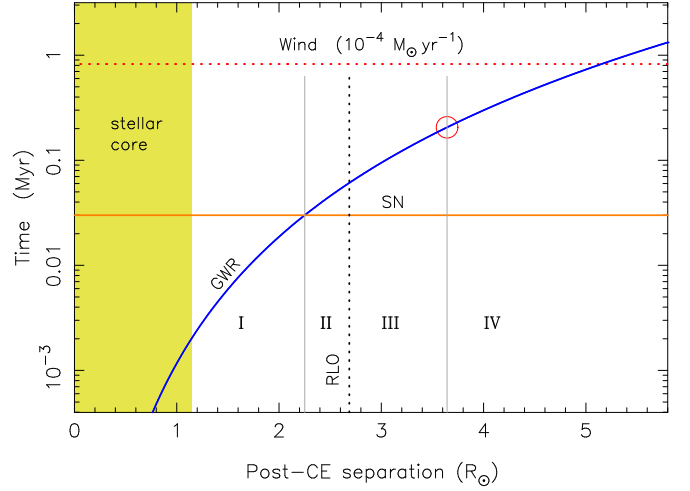}}
  \caption[]{
    Final fate of a post-CE binary system composed of a naked helium core and a BH as a function of orbital separation after envelope ejection.
    The chosen donor star model is that of the $88\;M_{\odot}$ star ($Z=Z_{\odot}/50$) at its maximum extent as a giant, cf. Figs.~\ref{fig:structure} (lower panel) and \ref{fig:structure_R}.
    The mass of the exposed helium core is $M=52.35\;M_{\odot}$ and the BH is assumed to have a mass of $30\;M_{\odot}$. 
    The blue line (GWR) represents the merger time due to gravitational-wave radiation, the orange line (SN) represents the remaining lifetime of the core until it collapses, and
    the red dotted line (Wind) represents the timescale ($a/\dot{a}_{\rm wind}$) of orbital widening due to stellar wind mass loss.
    The yellow shaded region marks the core region ($X_H<0.10$). See Sect.~\ref{subsec:preSN_merger}.
    }
\label{fig:timescale}
\end{center}
\end{figure}

For massive binaries undergoing CE evolution with deep in-spiral of the BH, we investigated if it is possible that the timescale of gravitational-wave radiation (GWR) 
of the post-CE binary is shorter than the remaining lifetime of the exposed core. The result is illustrated in Fig.~\ref{fig:timescale}.
We studied the fate of our $88\;M_{\odot}$ star ($Z=Z_{\odot}/50$) under the assumption of onset of a CE when this star is near its maximum extent as a giant 
($R=3530\;R_{\odot}$ at an age of 2.83~Myr, cf. Fig.~\ref{fig:structure} (lower panel) and Fig.~\ref{fig:structure_R}). 
Furthermore, we assumed a mass of the in-spiralling BH of $30\;M_{\odot}$.

In Fig.~\ref{fig:timescale} region~I marks the extreme case where GWR would be strong enough to merge the binary before a BH-BH binary is produced, that is, before
the collapse of the exposed core.
However, the vertical dotted line marks the orbital separation of the BH where the core would fill its Roche~lobe and continue mass transfer.
Hence, regions~I and II would most likely lead to an early merger in any case. In region~III, the orbit will decrease in size before core collapse as a result of GWR,
while in region~IV the orbit widens before core collapse as a
result of the stellar wind. 

For this system, we estimate the remaining lifetime of the post-CE exposed core ($M=52.35\;M_{\odot}$) to be about 30~kyr (given that the progenitor star was evolved close to the onset of carbon burning).
For the merger time of the binary we find from integration \citep[assuming time-independent point masses $M_1$ and $M_2$ in a circular orbit with separation, $a$, following][]{pet64}
\begin{equation}
 \tau _{\rm GRW} = \frac{1}{4}\,\frac{a}{|\dot{a}_{\rm GWR}|} = \frac{5}{256}\,\frac{c^5}{G^3}\,\frac{a^4}{M_1 M_2 (M_1+M_2)}\,,
 \label{eq:GWR}
\end{equation}
where $c$ is the speed of light.
The steep dependence on $a$ means that systems will spend most of their in-spiral time at a large separation. They only evolve to half their initial separation in 
about 94\% of the full merger time calculated from Eq.~(\ref{eq:GWR}).
Therefore, taking into account the finite size of the exposed core only changes the true merger time slightly. 

It is seen in Fig.~\ref{fig:timescale} that if a post-CE survival criterion is that the exposed core is not allowed to fill its Roche~lobe (i.e. ruling out all post-CE orbital separations to the
left of the vertical dotted line), then at least for this particular system, it is not possible for the post-CE system to merge as
a result of GWR before the
naked core terminates its life and undergoes core collapse to form a BH. Hence, all systems to the right of the dotted line would produce BH-BH binaries and eventually become LIGO sources.

We can determine the critical separation at which the orbital decay due to GWR ($\dot{a}_{\rm GWR}$) is exactly opposed by the orbital widening due to 
(fast, Jeans mode) stellar wind mass loss ($\dot{a}_{\rm wind}$) with a rate of $\dot{M}_{\rm wind}$ as
\begin{equation}
 a_{\rm crit} = \left( \frac{64}{5}\,\frac{G^3}{c^5}\,\frac{M_1 M_2 (M_1+M_2)^2}{\dot{M}_{\rm wind}}\right) ^{1/4}.
 \label{eq:breakpoint}
\end{equation}
For the system in Fig.~\ref{fig:timescale}, we find $a_{\rm crit}=3.65\;R_{\odot}$ (cf. red circle) for $\dot{M}_{\rm wind}=10^{-4}\;M_{\odot}\,{\rm yr}^{-1}$. 

\subsection{Extended envelopes of helium stars}\label{subsec:He-env}
Massive, luminous stars, both hydrogen-rich and helium (Wolf-Rayet) stars, reach the Eddington limit in their interiors 
and develop inflated and extremely diluted envelopes \citep{iuk99,ppl06,sglb15}.
Applying these extended radii for naked helium star models would prevent in-spiral to small separations 
if a criterion for survival of the CE ejection would be that the exposed helium core is not allowed
to fill its Roche~lobe. Hence, our estimated values of $R_{\rm core}$ (and thus $M_{\rm X,min}$) would be much higher
if such a conservative criterion was at work, which would make CE ejection even more difficult.

In our estimates of $R_{\rm core}$ for naked helium stars, we followed \citet{sglb15} and defined the
core radius as the distance from the centre of the star to where the ratio $\beta$ of gas pressure to total pressure
drops below 0.15 for the first time. This definition agrees fairly well with the location of the point
where the density gradient is steepest ($\partial^2 \log \rho / \partial m^2 = 0$).

To check the validity of this relaxed criterion for $R_{\rm core}$, we performed calculations of 
Roche-lobe overflow for a BH placed inside the extended envelope of a helium star, using the stellar evolution code BEC \citep[][and references therein]{ywl10}.
As expected, the BH simply peels off the outer envelope of the star, which might have been lost in a strong wind in any case.

To summarise, applying the relaxed criterion on $R_{\rm core}$ does not result in yet another episode of dynamically unstable mass transfer,
and by applying these smaller core radii, we therefore probe the conditions under which the CE is most easily ejected.

\subsection{Luminous blue variables}\label{subsec:LBV}
In our galaxy as well as in the Large Magellanic Cloud, stars more massive than $\sim\!60\;M_{\odot}$ are not found to be cooler than about 
20\,000\,K \citep{hd94,cfl+14}. That single stars in the considered luminosity range are thought to develop into hydrogen-poor 
Wolf-Rayet stars \citep{lhl+94,mm05} implies that they do loose their envelope even without the help from a binary companion. 
The so-called luminous blue variables (LBVs) are located close the this observational border \citep{svd04}, and the LBV variability and 
outbursts are thought to be connected to the stellar Eddington limit \citep{uf98,sglb15}.

The envelopes of stellar models near the Eddington limit may have very low binding energies \citep{gov12,sglb15}. In the limit of 
near zero binding energies, a companion star could indeed kick off these envelopes without the requirement of a significant in-spiral.  
A similar situation is reached in the final phases of the AGB evolution of low- and intermediate mass stars. For stars in this mass range, 
there is observational evidence that in some cases, the common envelope ejection occurs with an insignificant orbital decay \citep{nvyp00}. 

Consequently, stars near their Eddington limit, when they capture a companion into their envelope, may be prone to loose their envelope easily, 
but they will not produce sufficiently tight binaries to serve as progenitors for double compact mergers. The mapping of the Eddington limit 
throughout the parameter space of mass and metallicity is far from complete. \citet{uf98} pointed out that the Eddington limit is reached at 
higher masses for lower metallicity. This is confirmed by Sanyal~et~al. (2016, in~prep.), who find a limiting mass of 
$\sim\!100\;M_{\odot}$ at the metallicity of the Small Magellanic Cloud.

\subsection{Convective core overshooting}\label{subsec:overshoot}
The models presented in this work apply a convective core-overshooting parameter of $\delta_{\rm OV}=0.335$ pressure scale heights ($H_{\rm p}$), meaning that the radius of the convective core is equal to the radius given by the Ledoux criterion at the formal core boundary plus an extension equal to $0.335\,H_{\rm p}$.
Neglecting, or strongly reducing, the amount of convective core overshooting leads to a significantly different interior structure,
not only because of its reduced core mass, but also owing to the star burning the main part of its helium core already in the Hertzsprung gap,
before ascending the giant branch. For example, for our $20\;M_{\odot}$ model with $Z=Z_{\rm MW}$ at the base of the giant branch, the central helium mass abundance, 
$Y_c=0.24$ for $\delta_{\rm OV}=0.0$ compared to the case of $Y_c=0.99$ for $\delta_{\rm OV}=0.335$. 
The calculated envelope binding energies and $\lambda$-values are therefore also affected
by the choice of $\delta_{\rm OV}$. For example, for the $20\;M_{\odot}$ star mentioned above, we find that when it is evolved to a radius of $R=1200\;R_{\odot}$
, then $|E_{\rm bind}|$ can be almost a factor 10 smaller (and $\lambda$ a factor of 10 larger) using $\delta_{\rm OV}=0.0$ compared to $\delta_{\rm OV}=0.335$. 
The corresponding core masses are about $5.9\;M_{\odot}$ and $7.2\;M_{\odot}$, respectively. 
However, for the former case ($\delta_{\rm OV}=0.0$) the central mass density is significant higher, leading to less tightly bound envelopes.
For a $40\;M_{\odot}$ star we find that the impact of changing $\delta_{\rm OV}$ is smaller.

\section{Implications for LIGO detected BH-BH binaries}\label{sec:ligo}
\subsection{GW150914}\label{subsec:GW150914}
The two merging BHs in GW150914 were located at a redshift of about $z\simeq 0.09$ ($\sim\!400\;{\rm Mpc}$) and reported to have masses of 
$36^{+5}_{-4}\;M_{\odot}$ and $29^{+4}_{-4}\;M_{\odot}$ \citep{aaa+16}. 
These masses, as well as preliminary aLIGO detection rate estimates, agree well with the predictions of \citet{mlp+16} and \citet{dm16}.
Unfortunately, the spins of the individual BHs were not well constrained from this event.
The question is whether the CE formation channel can also reproduce an event like GW150914. 

Based on the analysis presented in this paper, we conclude that the CE formation channel might work (in a low-metallicity environment) to produce relatively 
massive BH-BH systems like GW150914 (which require $M_{\rm ZAMS}\ga 50\;M_{\odot}$). A caveat is that there are still many
uncertain aspects of CE ejection and that 3D hydrodynamical modelling in this direction is only at its infant stage, 
so far with no simulations of envelope ejections from a compact object embedded in the envelope of a massive star.

\subsection{GW151226}\label{subsec:GW151226}
GW151226 was reported to consist of a pair of BHs of masses $14^{+8}_{-4}\;M_{\odot}$ and $7.5^{+2.3}_{-2.3}\;M_{\odot}$ and is also located
at a redshift of $z\simeq 0.09$ \citep{aaa+16b}. 
It is notable that its total mass is about three times lower than the spectacular first event GW150914.
Thus the formation of GW151226 cannot be explained by the CHE/MOB scenario (which in addition to BH masses $\ga25\;M_{\odot}$ also predicts a mass ratio very close to unity).
From the simple energy budget analysis presented here, the CE formation channel could work for GW151226 in both a low- and high-metallicity 
environment (see Fig.~\ref{fig:Mxmin}), assuming that the $14\;M_{\odot}$ BH formed first. In the (somewhat unlikely) case that the $7.5\;M_{\odot}$ BH formed first, however,
an origin in a high-metallicity environment seems difficult. According to Fig.~\ref{fig:Mxmin}, we can see that such a BH can only remove the CE of a $Z=Z_{\rm MW}$ star
when the star has an initial mass lower than about $40\;M_{\odot}$ (and only when it is evolved to near its very maximum radial extent on the giant branch), which means that the mass of the 
collapsing core would be lower than about $18\;M_{\odot}$, according to our models. Hence, even modest mass loss of $\ga4\;M_{\odot}$ in the BH formation process would rule out this possibility,
depending on the exact masses of the two BHs.

\subsection{Comparison to other work}\label{subsec:belczynski}
In a recent paper by \citet{bhbo16}, a CE formation channel was put forward for GW150914.
While the various aspects of CE evolution discussed here in this paper are generic, comparing our results directly
with those of \citet{bhbo16} is difficult since few details of their applied stellar models are given. 
From their model (see their Fig.~1), we can deduce that the suggested $82.2\;M_{\odot}$ donor star ($Z\simeq Z_{\odot}/30$) has a radius of about
$1700\;R_{\odot}$ at the onset of the CE phase with a $35.1\;M_{\odot}$ BH accretor.
From our computed stellar structure models of an $80.0\;M_{\odot}$ star, we find $\lambda \sim\! 0.01$ \citep[in agreement with][]{dt01},
which yields $|E_{\rm env}|\simeq 5\times 10^{50}\;{\rm erg}$. However, after the in-spiral of the BH, the orbital
separation in the Belczynski~et~al. model is quoted to be $a_{\rm f}=43.8\;R_{\odot}$, which corresponds to a 
released orbital energy of $|E_{\rm orb}|\simeq 5.4\times 10^{49}\;{\rm erg}$, that is, about 10~times too small to eject the envelope.
However, such an apparent discrepancy could be an artefact of simply applying different convective core-overshooting parameters,
and given the relatively low stellar core masses in their illustrated scenario, we suspect that \citet{bhbo16} applied stellar models with
small convective core overshooting. Alternatively, it is possible that they included released accretion energy as a main energy source in their budget, 
although this aspect is not discussed in their paper.

\subsection{BH-BH formation: stability of the first RLO}\label{subsec:std_RLO}
Following the CE scenario for producing BH-BH binaries, we can also make predictions for the dynamical stability of the first mass-transfer phase (RLO).
To lower the binding energy of the envelope during the CE phase and thus enhance the chance for surviving the in-spiral of the BH,
the pre-CE binary system must be wide to secure an evolved donor star (Figs.~\ref{fig:Ebind} and \ref{fig:Mxmin}). To fulfil the requirement of a wide pre-CE system, this means that
the first mass-transfer phase from the primary star (the progenitor of the first-formed BH) to the (less evolved) secondary star must be dynamically stable 
or, in case of unstable RLO (see also Sect.~\ref{subsec:LBV}), the orbital separation is only slightly reduced. 
Otherwise, if this first mass-transfer phase would form an effective CE, it would either reduce the orbital size drastically or result in an early coalescence, thus
preventing the subsequent formation of a BH-BH system.
However, the stability of mass transfer in massive binaries with non-degenerate stars is largely unexplored in the literature. 
In particular, we question to which extent binary systems would be dynamically stable at this stage
since the timescale of the mass transfer is often much shorter than the thermal timescale of the accreting star.
As a possible result, the accreting star may expand, initiate mass loss through the second Lagrangian point and result in a CE.
It cannot be ruled out therefore that a significant fraction of the systems would merge already in this early phase. However, further investigations are needed in this direction
before any conclusion can be drawn.

\section{Conclusions}\label{sec:conclusions}
We have analysed the CE ejection process in post-HMXB systems. 
From our investigation of stellar structures and energy budget considerations, we find that CE evolution, in addition to producing double WD and double NS binaries, 
may in principle also produce massive BH-BH systems with individual BH component masses of up to $60\;M_{\odot}$ \citep[beyond which point a pair-instability SN is expected
to lead to complete disruption of the progenitor star and not leaving behind any compact remnant, cf.][]{hw02,cw12}. The potential for successful CE ejection
is particularly good for donor stars evolved to giants beyond the Hertzsprung gap. 

We find that the change in the $\lambda$-parameter with stellar radius is significantly more important than changes caused by different stellar masses or metallicities
(although some mass dependence is noted on the giant branch).
The associated binding energies of the stellar envelopes increase with stellar mass (independent of evolutionary status), but are generally independent of metallicity
(except for massive high-metallicity stars that evolve to become LBV stars).
The convective core-overshooting parameter applied in stellar models, $\delta_{\rm OV}$, however, can strongly affect the calculated values of $\lambda$ and $E_{\rm bind}$
(up to a factor of 10). 

Based on our detailed analysis of the evolution of the interior structure of massive stars, it is evident that the difficulty in determining the precise 
bifurcation point (core boundary) remains the key uncertain aspect of the outcome of CE evolution \citep{td01}.
Whereas the hydrogen abundance \citep[$X_{\rm H}=0.10$,][]{dt00} and the maximum-compression point criteria \citep{iva11} roughly yield similar locations for the core boundary, 
it remains uncertain if the in-spiral continues significantly below the bottom of the convection zone in the envelope.  
Until future 3D hydrodynamical simulations will succeed in ejecting the CE of massive stars, the estimated LIGO detection rates from population synthesis
of merging BH-BH and NS-NS binaries \citep{aaa+10} will remain highly uncertain, not to mention all other aspects of binary evolution and interactions not investigated in this work. 

We explored the importance of additional energy sources to help ejecting the CE. We confirm that recombination energy makes an important contribution to
the total internal energy in low- and intermediate-mass stars. However, for massive ($>30\;M_{\odot}$) stars the contribution may be less than 1\%, depending on the core boundary.
Hence, liberated recombination energy may not play any significant role in forming BH-BH binaries through CE evolution. 

The release of accretion energy, on the other hand, from an in-spiralling compact object (BH or NS), can be significant for the energy budget
and may help to facilitate the CE ejection process. However, models of time-dependent energy transport in the convective envelope are needed to quantify this effect. 

Although a deep in-spiral of a BH is possible in massive binaries that may eject their envelope, the exposed core will most likely terminate
its evolution, and collapse before GWR would cause such post-CE binaries to coalesce.
Hence, once a post-CE system is formed composed of a helium (Wolf-Rayet) star and a BH, the outcome is expected to be a BH-BH binary. 

While it is difficult to estimate the outcome of CE evolution with high confidence, the arguments presented in this paper taken together 
suggest that it seems realistic to expect that production of BH-BH binaries are possible through the CE\ formation channel, leading to events such as GW150914 and GW151226.

\begin{acknowledgements}
MUK acknowledges financial support by the DFG Grant: TA 964/1-1 awarded to TMT. D.Sz. was supported by GA\v{C}R grant 14-02385S.
\end{acknowledgements}

\bibliographystyle{aa}
\bibliography{tauris_refs}

\end{document}